\documentclass[11pt,english]{article}
\usepackage[dvips]{graphicx}
\usepackage{amssymb, mathrsfs, amsmath, setspace, fancyhdr, xcolor, ifpdf, graphicx, rotating, comment, tikz, url, float, cite}
\definecolor{darkblue}{rgb}{0,0,0.6}
\usepackage[colorlinks=true,citecolor=darkblue,linkcolor=black,urlcolor=darkblue]{hyperref}
\usepackage[font=small,labelfont=bf]{caption}
\usepackage[utf8]{inputenc}

\newcommand{\f}{\frac}

\def\be{\begin{eqnarray}}
\def\ee{\end{eqnarray}}

\usepackage{amsmath,amsfonts,amssymb,bm}
\usepackage{color}

\begin{document}
\unitlength = 1mm
\ 

\begin{center}

{ \LARGE {\textsc{\begin{center}Sparseness bounds on local operators in holographic CFT$_d$ \end{center}}}}
\vspace*{1.7cm}
Eric Mefford,$^1$ Edgar Shaghoulian,$^{1,2}$ and Milind Shyani$^3$\\
\vspace*{0.6cm}
\small{
$^1$Department of Physics, University of California, Santa Barbara, CA 93106 USA\\
$^2$Department of Physics, Cornell University, Ithaca, NY 14853 USA\\
$^3$Stanford Institute for Theoretical Physics, Stanford, CA 94305, USA\\}
\vspace{0.5cm}

\vspace*{0.6cm}

\end{center}

\begin{abstract}

\noindent We use the thermodynamics of anti-de Sitter gravity to derive sparseness bounds on the spectrum of local operators in holographic conformal field theories. The simplest such bound is $\rho(\Delta) \lesssim \exp\left(\f{2\pi\Delta}{d-1}\right)$ for CFT$_d$. Unlike the case of $d=2$, this bound is strong enough to rule out weakly coupled holographic theories. We generalize the bound to include spins $J_i$ and $U(1)$ charge $Q$, obtaining bounds on $\rho(\Delta, J_i, Q)$ in $d=3$ through $6$. All bounds are saturated by black holes at the Hawking-Page transition and vanish beyond the corresponding BPS bound. 
\end{abstract}

\pagebreak
\setcounter{page}{1}
\pagestyle{plain}

\setcounter{tocdepth}{1}
\section{Introduction}
There has been a recent surge of interest in precisely characterizing  conformal field theories with a weakly coupled Einstein gravity dual, with equations now accompanying folkore from the past. The most quantitative work has focused on conformal field theories in two dimensions
, though there has also been progress on higher-dimensional theories
. The difficulties brought on by higher dimensions are clear: the constraining infinite-dimensional Virasoro symmetry is absent and modular invariance of the torus partition function does not immediately provide constraints on the space of local operators.

In this paper we will use the familiar thermodynamics of gravity in asymptotically anti-de Sitter spacetimes to provide quantitative sparseness bounds on the spectrum of local operators of holographic conformal field theories. This approach began with \cite{Hartman:2014oaa}, which showed that the thermodynamics of gravity in AdS$_3$ is reproduced if and only if the spectrum of operators with scaling dimension $\Delta < c/6$ and $\Delta \sim O(c)$ obeys $\rho(\Delta)\lesssim \exp(2\pi \Delta)$. This methodology was subsequently generalized to supersymmetric theories \cite{Benjamin:2015hsa}, correlation functions \cite{Kraus:2017kyl}, and higher-dimensional theories on tori \cite{Belin:2016yll}. The universality of the thermodynamics for holographic CFTs on tori can also be derived from the special center symmetry structure of such theories through the Eguchi-Kawai mechanism \cite{Shaghoulian:2016xbx}.

In two dimensions, the low-temperature and high-temperature thermodynamics are related to one another by modular invariance. This is what allows one to capture the entire thermodynamic phase structure by constraining only the low-lying ($\Delta < c/6$) operators. Unfortunately, in higher-dimensional theories on $S^{d-1}$, there is no obvious high-temperature/low-temperature duality. But there is still a universal feature of the gravitational phase structure that we can aim to reproduce from the CFT: the Hawking-Page phase structure \cite{Hawking:1982dh}, where, as a function of some external chemical potentials, the vacuum-subtracted free energy (or the entropy) jumps from $O(1)$ to $O(N^k)$ for $k$ some positive number. (For notational simplicity we will ignore the possibility of intermediate scalings $O(N^m)$ for $0<m<k$.) 
More specifically, we will reproduce the fact that the theory is confined ($O(1)$ scaling in the entropy) below the Hawking-Page transition temperature $T_{HP}$. 

To illustrate the basic idea, consider the finite-temperature canonical ensemble with normalization $E_{\text{vac}}=0$ and a deconfining phase transition at $\beta_c\sim O(1)$. Then,
\begin{align}
\log Z(\beta) \sim  \begin{cases}
\mathcal{O}(1),\qquad &\beta>\beta_c\\
\mathcal{O}(N^k),\qquad &\beta<\beta_c\,.
\end{cases}
\end{align}
Since $Z(\beta) = \int e^{-\beta \Delta}\rho(\Delta) d\Delta$, the $O(1)$ behavior of $\log Z$ may be ruined if the density of states $\rho(\Delta)$ grows too quickly for states with $\Delta \gtrsim O(N^k)$. More precisely, we have
\be
\log Z(\beta > \beta_c) \sim O(1)\qquad \text{if and only if} \qquad \rho(\Delta) \lesssim e^{\beta_c \Delta} \quad \text{for}\quad  \Delta \gtrsim O(N^k)\,.
\ee
In the worst-case scenario where the bound is saturated for all states, we have
\be
 Z(\beta>\beta_{c}) =  \int_0^\infty d\Delta e^{-\beta \Delta} e^{\beta_{c}\Delta} =\f{1}{\beta - \beta_{c}} \,.
\ee
Hence, $\log Z$ is $O(1)$ for all $\beta>\beta_c + \epsilon$ for $\epsilon \ll 1$ as long as $\epsilon$ is not exponentially small in $N$. 


While a deconfinement transition is generically expected for large-$N$ adjoint CFTs on compact spaces \cite{Witten:1998zw, Aharony:2003sx}, it is the precise temperature at which the transition occurs which gives us mileage. In particular, applying the above argument to the well known Hawking-Page transition at inverse temperature $\beta_{HP}=\f{2\pi}{d-1}$ gives us a bound on the spectrum of local operators of holographic CFTs:
\be\label{vacuumbound}
\rho(\Delta) \lesssim \exp\left(\f{2\pi \Delta}{d-1}\right)\quad \text{for}\quad  \Delta \gtrsim O(N^k) \qquad \text{if and only if} \qquad \log Z(\beta > \beta_{HP}) \sim O(1) \,.
\ee
This bound applies to the entire spectrum, but above the transition temperature, bulk thermodynamics tells us that the large-$N$ density of states is given by the degeneracy of the black hole dominating the ensemble, which is generically smaller than our bound (see figure \ref{boundvsentropy}). Interestingly, our bound must be saturated at the transition point, since at leading order in $N$ we can write
\be
F_c= E_c- S_c/\beta_c = 0 \implies S_c = \beta_c E_c \implies \rho(E_c) = e^{\beta_c E_c}\,,
\ee
where we are assuming that immediately above the transition we have equivalence of canonical and microcanonical ensembles, i.e. $E_c \equiv \langle E \rangle_{\beta_c}$ is a well-defined energy level stable to fluctuations. Applied to AdS/CFT, this argument means that our bound will be saturated by the black hole at the Hawking-Page transition. In appendix \ref{bhentropy}, we invert the logic behind this fact to provide a field-theoretic density of states interpretation for the Bekenstein-Hawking entropy. 

For the remainder of this paper, we generalize eq. (\ref{vacuumbound}) using known classical black hole solutions to bound the density of operators of the dual CFT with given scaling dimension $\Delta$, spins $J_i$ and $U(1)$ charge $Q$. For $d=2$ the bounds will reduce to those of \cite{Hartman:2014oaa}. Importantly, these bounds are more constraining in $d>2$ than in $d=2$, because for $d=2$ modular invariance implies that, if a single deconfining phase transition occurs, it must occur at $\beta = 2\pi$ independent of coupling. Indeed, free symmetric orbifolds (which are not dual to weakly coupled Einstein gravity theories) have a transition at $\beta = 2\pi$ just like AdS$_3$ gravity, and $\alpha'$ perturbation theory around AdS$_3$ gravity leaves the Hawking-Page temperature unchanged \cite{Kraus:2005vz}. On the other hand, in higher dimensions the deconfining temperature tends to increase as interactions are turned on. For example, in both ABJM theory and $\mathcal{N}=4$ super Yang-Mills, it can be checked that $\beta_{HP}(\lambda = 0) > \beta_{HP}(\lambda = \infty)$  for 't Hooft coupling $\lambda$ \cite{Sundborg:1999ue, Aharony:2003sx, Nishioka:2008gz}, with further calculations suggesting monotonic behavior between the free and strongly coupled theories \cite{Landsteiner:1999gb, Spradlin:2004pp, Papathanasiou:2009en}. This means that $\log Z \sim O(1)$ for a smaller range of temperatures as the interaction strength is decreased. By the argument above, this means that weakly coupled CFTs must be less sparse---they \emph{must} have $\rho(\Delta) \gtrsim e^{2\pi\Delta/(d-1)}$ somewhere in their spectrum. The fact that strong interactions are necessary to reproduce the precise low-temperature phase structure of AdS gravity in higher dimensions has been translated into a simple bound on the density of local operators. The violation of our bound is a sharp diagnostic of ``how much" interactions have to sparsify a spectrum. There is another interesting aspect to these bounds that we will discuss in section \ref{bpssec}: they imply an $O(1)$ density of states beyond corresponding BPS/unitarity bounds. For example, taking $\Delta <0$ implies $\log \rho(\Delta) \sim \mathcal{O}(1)$, which looks like a coarse unitarity bound. 

The layout of the rest of the paper is as follows. In section \ref{methodsec}, we provide the methodology behind obtaining our bounds more carefully. In section \ref{genericsec}, we provide calculational details for deriving our various bounds. Analytic bounds are possible for three parameters, either mass and two spins or mass, one spin and one $U(1)$ charge, but for four or more parameters, we must resort to numerics. Two-parameter analytic results are summarized in table \ref{analyticdensities}. In section \ref{bpssec}, we discuss the connection of our bounds to BPS/cosmic censorship bounds. In section \ref{cvsec}, we speculate on the connection between the high-lying spectrum or high-temperature thermodynamics and our bounds on the low-lying spectrum. We will begin with an analysis of the Cardy-Verlinde formula, which correctly gives the entropy above the Hawking-Page temperature $T_{HP}$ for holographic CFT$_d$ on $S^{d-1}$ \cite{Verlinde:2000wg}. After discussing the many limitations of this formula, we instead focus on a more robust feature of the high-temperature thermodynamics: the extended range of validity of a high-temperature effective field theory. In appendix \ref{bhentropy}, we provide a field-theoretic density of states interpretation for the Bekenstein-Hawking entropy of black holes at Hawking-Page phase transitions. In appendix \ref{higherdetails}, we provide details for calculations in $4\leq d \leq 6$.  


\section{Method for obtaining bounds}\label{methodsec}
In this section, we explain more carefully our method for obtaining bounds on the allowed density of states of operators with $U(1)$ charge and spin for holographic theories with a confining phase transition. We consider a grand canonical ensemble at finite temperature $\beta$, with $m$ angular velocities $\Omega_i$ and a single chemical potential for $U(1)$ charge $\Phi$ for CFT$_d$:
\be
\log Z(\beta, \Omega_i, \Phi)= \int dE\, dJ_i \,dQ \exp\left[-\beta (E -\Omega_i J_i-\Phi Q)\right]\rho(E, J_i, Q)\,,
\ee
where the integral goes over the spectrum of the theory and we sum over repeated indices in the exponential. Except when otherwise noted, we will always normalize the ground state energy (even for $d=2$) to zero. The extension to additional chemical potentials is trivial.

A confining phase transition means that $\log Z[\beta>\beta_c(\Omega_i, \Phi)] \sim O(1)$, i.e. the free energy does not scale with $N$ for temperatures below some critical temperature $\beta_c^{-1}(\Omega_i, \Phi)$. The chemical potentials $\Omega_i,\, \Phi$ and $\beta$ span an $(m+2)$-dimensional space, and the confinement-deconfinement phase transition happens on a co-dimension one critical surface $\beta=\beta_c(\Omega_i,\Phi)$. The $O(1)$ scaling of the free energy requires that the density of states be bounded from above,
\begin{align}
\rho(E,J_i,Q)\lesssim \exp\left[\beta \left(E-\Omega_i J_i -\Phi Q \right)\right]\,,\qquad \forall\quad \beta, \Omega_i, \Phi \text{ in the confined phase.}\,
\label{boundform}
\end{align}
It is simple to minimize the right-hand-side with respect to the potentials $\beta$, $\Omega_i$, and $\Phi$ to provide the tightest bound.
In the case of $\Omega_i = \Phi= 0$ the minimization gives $\beta = \beta_c$ for $E> 0$ and the bound becomes $\rho(E)\lesssim e^{\beta_c E}$, while for $E<0$ gives $\beta \rightarrow \infty$ and our bound vanishes. 
This behavior is generic: the minimum of eq. \eqref{boundform} always lies either on the critical surface or at $\beta \rightarrow \infty$ which gives vanishing degeneracy. The set of values for charges which separates the two behaviors corresponds to a unitarity/BPS bound. To see the two behaviors in general, we first impose parity symmetry under $\Omega_i \rightarrow -\Omega_i$ such that the critical surface is an even function of the chemical potentials $\Omega_i$ and $\Phi$. Since eq. \eqref{boundform} is invariant under $\{J_i,Q_i,\Omega_i,\Phi \} \rightarrow \{-J_i,-Q_i,-\Omega_i,-\Phi\}$, it is then sufficient to consider only operators with $\{J_i,Q\}>0$ and potentials with $\{\Omega_i,\Phi\} > 0$. For the theories we consider, these potentials $\Omega_i, \Phi$ have finite range, being bounded below by $\Omega_i=0$ and $\Phi=0$ and above by some constants which depend on the theory and dimension. Since $\beta$ is an overall multiplicative factor, we can minimize it independently, landing on $\beta_c(\Omega_i, \Phi)$ if $E-\Omega_i J_i - \Phi Q>0$ for all $\{\Omega_i,\Phi\}$ and $\beta\rightarrow \infty$ otherwise. In the former case we then minimize along the critical surface, while in the latter case the bound simply vanishes. 
The minimization along the critical surface is 
\begin{align}
\nabla\left[\beta_c\left(E-\Omega_i J_i-\Phi Q\right)\right] = 0
\end{align}
for a given set of charges $\{E, J_i,Q\}$, and $\nabla = (\partial/\partial\Omega_1,\dots,\partial/\partial\Omega_m,\partial/\partial\Phi)$. 

Until this point, the discussion applies to states with general $U(1)$ charge $Q$ and momenta $J_i$ in large-$N$ gauge theories with a confining phase transition. Focusing on local operators in holographic CFTs with a semiclassical Einstein gravitational dual, we restrict to dimensions $ 2\leq d  \leq 6$ and the spatial manifold $S^{d-1}$. The Hawking-Page temperature in the bulk will serve as the deconfinement temperature in the CFT. 
To find the Hawking-Page transition, we compare the on-shell action of the relevant black hole solution to that of vacuum AdS. The vacuum AdS solution will have topological identifications and constant gauge field to match the inverse temperature, angular velocities, and chemical potential for $U(1)$ charge of the black hole. When the black hole has charge and spin, the deconfinement temperature will depend on the chemical potential $\Phi$ and angular velocities $\Omega_j$. Below this temperature, the dual CFT is in a confined phase (dual to the AdS vacuum) and above this temperature the dual CFT is deconfined (dual to a black hole).

We consider the most general black holes in $d+1$ dimensions for the cases $d=2$ through $d=6$ with a single $U(1)$ charge and  $\left\lfloor\frac{d}{2} \right\rfloor$ spins.  These black holes are asymptotic to a (spinning) Einstein static universe (ESU) which, in the Lorentzian case, has topology $\mathbb{R}\times S^{d-1}$. Classical solutions for the generically spinning charged black hole in dimensions $d=5$ and $d=6$ depend on choice of supergravity truncation and so our results in those cases should be considered in that context. Nevertheless, bounds obtained from these solutions are similar to their lower dimensional counterparts. Analytic results are possible in all dimensions for up to three parameters, while numerics are necessary for four and five parameters. Two-parameter bounds are shown in table \ref{analyticdensities}. Analytic expressions are only applicable when they are real and positive; when they become complex or negative it means the charges admit a set of chemical potentials for which $E-\Omega_i J_i - \Phi Q<0$ and the minimization procedure lands at $\beta \rightarrow \infty$ instead of the Hawking-Page surface. This leads to an $O(1)$ density of states. 
\begin{table}[t!]
\begin{center}
{\renewcommand{\arraystretch}{2.5}
 \begin{tabular}{|c | c | c | c |} 
 \hline
$d$ & $\log \rho(\Delta)$ & $\log \rho(\Delta,Q)$ & $\log \rho(\Delta,J)$ \\ [1ex] 
 \hline\hline
 2 & $2\pi \Delta$ & N/A & $2\pi \Delta\sqrt{1-J^2/\Delta^2}$ \\ 
 \hline
 3 & $\pi \Delta$ & $\pi \Delta\sqrt{1-Q^2/\Delta^2}$ & $\pi \Delta(1-J^2/\Delta^2)$ \\ 
 \hline
 4 & $\frac{2\pi \Delta}{3}$ & $\frac{2\pi \Delta}{3}\sqrt{1-\frac{3}{4}Q^2/\Delta^2}$ & $\frac{2\pi\Delta}{3} \left(2-\sqrt{1+3 J^2/\Delta^2}\right)$ \\
 \hline
 5 & $\frac{\pi \Delta}{2}$ & $\frac{\pi \Delta}{2}\sqrt{1-\frac{2}{3}Q^2/\Delta^2}$ & $\frac{\pi\Delta}{4}  \left(3-\sqrt{1+8 J^2/\Delta^2}\right)$  \\
 \hline
 6 & $\frac{2\pi \Delta}{5}$ & $\frac{2\pi \Delta}{5}\sqrt{1-\frac{5}{8}Q^2/\Delta^2}$ & $\frac{2\pi\Delta}{15} \left(4-\sqrt{1+15 J^2/\Delta^2}\right)$ \\ [3pt]
 \hline
\end{tabular}}
\caption{\label{analyticdensities} Bounds on the density of states for charged spinless operators (second column) and uncharged spinning operators (third column). When these expressions become complex or negative, the bound instead is $\log \rho = 0$.
}
\end{center}
\end{table}

Notable in this table is the absence of a bound for operators with $U(1)$ charge in 2d CFTs. Electrically charged static black holes in three dimensions have interesting but somewhat peculiar thermodynamic properties---see \cite{Jensen:2010em, Kraus:2006wn}. Among these properties is the fact that if one wants to include a bulk Maxwell field, the black hole mass is not bounded from below \cite{Martinez:1999qi}. If one wants to consider only a Chern-Simons term -- which is necessary to describe a $U(1)$ current on the boundary -- there are new difficulties in finding the dominant saddle. 
It is unclear how to match asymptotics as any non-zero holonomy of the gauge field remains constant along the radial direction. A holonomy in the spatial direction would lead to a singularity at the origin for the vacuum AdS phase, while a holonomy in the thermal direction would lead to a singularity at the horizon for the black hole phase. If one includes both Maxwell and Chern-Simons terms for the same $U(1)$ gauge field, the spacetimes include closed timelike curves in the asymptotic region \cite{Banados:2005da}.  Thus we cannot consistently analyze this situation in Einstein gravity coupled to $U(1)$ Chern-Simons and/or Maxwell gauge fields.

\section{Bounds on operators}\label{genericsec}
In this section, we derive our bounds for electrically charged operators with spin in CFT dimension $d=2$ through $d=6$. 
We begin with $d=3$ in section \ref{d3bound}, giving all details of the derivation of the bound. For general $d$ we state our analytic results, without derivation, for single-charge spinless operators in section \ref{charge}, single-spin uncharged operators in section \ref{spin}, double-spin uncharged operators in section \ref{spinspin}, and single-spin single-charged operators in section \ref{spincharge}. 

In the case with four or more parameters, we do not have an analytic bound but present numerical results in \ref{numal}. Figures for our numerical results will be presented together at the end of this section to emphasize the similarities between dimensions. 
The bound on the density of states  decreases when charge or spin is added, to the point that no states are allowed beyond a curve that exactly coincides with the BPS bound. As we will see, when the parameters satisfy the BPS condition and admit a BPS black hole, our bound is saturated by the entropy of the BPS black hole, 
\begin{align}
S_{BH} = max\left[\log \rho(\Delta_{BPS},Q_{BPS},J_{BPS,i})\right].
\end{align}
This is a special case of the fact that generic black holes at the Hawking-Page transition have an entropy which saturates our bound. 

\subsection{Example: $\rho(\Delta,Q,J)$ in $d=3$.}\label{d3bound}

In $d=3$, the AdS-Kerr-Newman black hole is the generic electrically charged, spinning black hole with AdS$_4$ asymptotics. Its thermodynamics were first studied in \cite{Caldarelli:1999xj}. In the limit of zero spin, the thermodynamics reproduces \cite{Chamblin:1999hg, Chamblin:1999tk}, and in the limit of zero charge reproduces \cite{Hawking:1998kw, Gibbons:2004uw, Gibbons:2004js}. The metric may be written
\begin{align}
ds^2 = -\frac{\Delta_r}{\rho^2}\left[dt-\frac{a\sin^2\theta}{\Xi}d\phi\right]^2 + \frac{\rho^2}{\Delta_r}dr^2+\frac{\rho^2}{\Delta_\theta}d\theta^2+\frac{\Delta_\theta\sin^2\theta}{\rho^2}\left[adt-\frac{r^2+a^2}{\Xi}d\phi\right]^2,
\label{4dKN}
\end{align}
where the metric functions and Maxwell field, $A$, are
\begin{align}
\Delta_r&=(r^2+a^2)(1+r^2)-2mr+q^2, \quad\quad\quad\quad\quad\quad\quad\quad \Delta_\theta=1-a^2\cos^2\theta\nonumber\\
\rho^2&=r^2+a^2\cos^2\theta, \quad\quad\quad\quad\quad\Xi=1-a^2, \quad\quad\quad\quad\quad A = -\frac{qr}{\rho^2}\left(dt-\frac{a\sin^2\theta}{\Xi}d\phi\right).
\end{align}
The mass $M$, angular momentum $J$, and electric charge $Q$---calculated via boundary integrals---are
\begin{align}
M = \frac{m}{G\Xi^2},\qquad 
J = 
\frac{am}{G\Xi^2}, \qquad
Q= \frac{1}{8\pi G}\int_{S^{d-1}_\infty} \star F=\frac{q}{G\Xi}.
\end{align}

\noindent Note that we follow the convention of \cite{Kostelecky:1995ei} for the normalization of Killing vectors as the associated conserved charges generate the $SO(d,2)$ algebra. To find the on-shell Euclidean action, we evaluate
\begin{align}
\label{IE4}
I_E = \frac{1}{16\pi G}\int d^4x \sqrt{g}(6+F^2) - \frac{1}{8\pi G}\int_{r=\Lambda} d^3x \sqrt{\gamma} K + \frac{1}{8\pi G}\int_{r=\Lambda} d^3x\sqrt{\gamma}\left(2+\frac{1}{2}R[\gamma]\right).
\end{align}

\noindent The second term is the Gibbons-Hawking-York boundary term and the last term is a local boundary counterterm that regularizes the action \cite{Balasubramanian:1999re}. 
The horizon angular velocity and inverse Hawking temperature of these black holes are
\begin{align}
\Omega_h = \frac{\Xi a}{r_+^2+a^2},\quad\beta=\frac{4\pi r_+(r_+^2+a^2)}{r_+^2(1+a^2)+3r_+^4-(a^2+q^2)}.
\label{4domega}
\end{align} 
The appropriate thermodynamic potential for spin, however, is the difference between $\Omega_h$ and $\Omega_\infty$, the angular velocity of the boundary ESU. One way to find this $\Omega_\infty$ is to boost the boundary metric to a static frame through a coordinate change $T= t-\Omega_\infty\phi$, giving $\Omega_\infty = -a$. We then obtain,
\begin{align}\label{4dom}
\Omega = \Omega_h - \Omega_\infty = \frac{a(1+r_+^2)}{r_+^2+a^2}.
\end{align}
The parameter $\Phi$ is chosen so that the gauge potential vanishes on the outer horizon, defined by $\Delta_r(r_+)=0$. Notably, this is the potential difference between the horizon and the conformal boundary, and serves as a chemical potential for $U(1)$ charged operators in the CFT.
\begin{align}
\Phi \equiv A_ak^a\biggr|_{r\to\infty}-A_ak^a\biggr|_{r=r_+}= \frac{qr_+}{r_+^2+a^2},
\label{4dphi}
\end{align}
where $k=\partial_t + \Omega_H \partial_\phi$ is the null generator of the horizon.
Subtracting the vacuum AdS result from the AdS-Kerr-Newman result gives
\begin{align}
\Delta I_E = \frac{\beta}{4G\Xi r_+}\left[(a^2+r_+^2)(1-r_+^2)+q^2\frac{a^2-r_+^2}{r_+^2+a^2}\right].
\end{align}
We can replace $\{r_+, a, q\}$ with $\{\beta, \Omega, \Phi\}$ using eq. \eqref{4domega} and \eqref{4dphi}. At fixed $\{\beta, \Omega, \Phi\}$, there are two competing stable phases--a large AdS-Kerr-Newman black hole and vacuum AdS. The bulk undergoes a Hawking-Page phase transition when the two saddle point solutions exchange dominance, in other words when $\Delta I_E=0$. In the limit of zero charge, the Hawking-Page transition occurs at $r_+=1$. In the limit of zero angular momentum, the Hawking-Page transition occurs at $r_+ = \sqrt{1-\Phi^2}$. For non-zero charge and angular momentum, it is simplest to extremize 
\begin{align}
\beta_{HP}(\Omega,r_+)(\Delta-\Omega J - \Phi_{HP}(\Omega,r_+) Q)
\label{extremum4d}
\end{align}
with respect to $\Omega$ and $r_+$. Obtaining the critical values for  $\Omega$ and $r_+$, we find that
\begin{align}
\label{4danalyticbound}
\log \rho(\Delta, J, Q) \lesssim \frac{\pi  \Delta}{\sqrt{2}}  \sqrt{\left(1+\hat{J}^2\right) \left(1+\hat{J}^2-\hat{Q}^2\right)+\left(1-\hat{J}^2\right) \sqrt{\left(1+\hat{J}^2-\hat{Q}^2\right)^2-4 \hat{J}^2}-4 \hat{J}^2},
\end{align}
where $\hat{J}=J/\Delta,\, \hat{Q}=Q/\Delta$. 
Note that if $\hat{J}+\hat{Q}>1$, eq. (\ref{4danalyticbound}) breaks down and the correct minimization gives an $O(1)$ density of states. This limit corresponds to the BPS bound $\Delta=|J|+|Q|$ for the lightest charged, spinning state. Notably, at $\Delta=|J|+|Q|$, the upper bound on our density of states exactly matches the degeneracy of the corresponding BPS black hole with those charges,
\begin{align}
S_{BH} = \text{max}\left[\log \rho(\Delta,\pm(\Delta-|Q|),Q)\right] = \pi Q \sqrt{1-Q/\Delta}.
\end{align}
\begin{figure}[t!]
\begin{center}
\includegraphics[scale=.55]{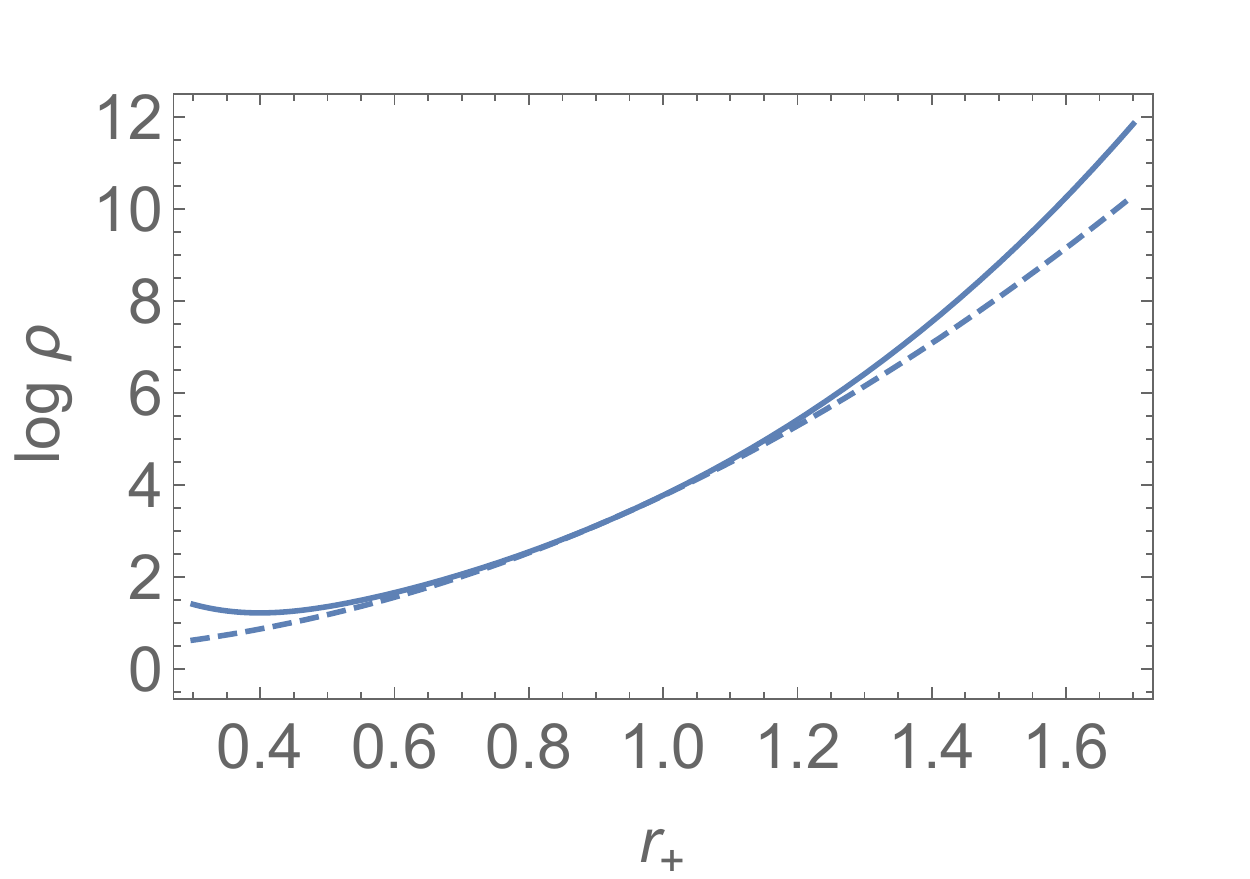}
\caption{\label{boundvsentropy} As an example, we plot our bound $\log \rho(\Delta, J, Q)$ (thick) and the entropy of the corresponding black hole (dashed) in $d=3$ for fixed black hole parameters, $a=.3, q=.4$. The two coincide at $r_{HP}$ and our bound is otherwise larger.}
\end{center}
\end{figure}Again we see that the upper bound on the density of states is saturated by the degeneracy of the bulk black hole at the Hawking-Page transition and is greater for all other black holes (see figure \ref{boundvsentropy}). For $\Delta=|J|+|Q|$, in $d=3$, the black hole at the phase transition is a BPS black hole.


\subsection{Charged, spinless operators}\label{charge}
\label{chargedanalytic}
To bound the density of states of charged, spinless operators, we examine the thermodynamics of ($d+1$)-dimensional AdS-Reissner-Nordstr\"om black holes. Using the conventions of \cite{Chamblin:1999tk}, the mass, global $U(1)$ charge, $U(1)$ potential, and inverse temperature for this black hole are
\begin{align}
\label{massandchargeRN}
M = \frac{(d-1)\omega_{d-1}}{16\pi G}m,\quad Q =\frac{(d-1)\omega_{d-1}}{8\pi G}cq, \quad \Phi = \frac{1}{c}\frac{q}{r_+^{d-2}}, \quad \beta = \frac{4\pi r_+^{2d-3}}{dr_+^{2(d-1)}+(d-2)[r_+^{2(d-2)}-q^2]},\nonumber
\end{align}
where $\omega_{d-1}$ is the area of the unit $(d-1)$ sphere, and $c=\sqrt{2(d-2)/(d-1)}$. The vacuum subtracted Euclidean action is
\begin{align}
\Delta I_E = \frac{\omega_{d-1}\beta}{16\pi G}\left[(1-c^2\Phi^2)-r_+^2\right]r_+^{d-2}.
\end{align}

\noindent As before, there are two competing stable phases at fixed $\Phi, \beta$. The first is the AdS vacuum with $m=q=0$ and constant gauge potential and the second is a large black hole, both at inverse temperature $\beta$. Solving for $\Delta I_E=0$, it is clear that for $r_{+} > \sqrt{1-c^2\Phi^2},$ black holes dominate the grand canonical ensemble while the vacuum dominates below. The corresponding Hawking-Page temperature is
\begin{align}
\beta_{HP}(\Phi)\biggr|_{r_+=\sqrt{1-c^2\Phi^2}} = \frac{2\pi}{(d-1)\sqrt{1-c^2\Phi^2}}.
\end{align}
Interestingly, for $\Phi = 1/c$, the Hawking-Page temperature $1/\beta_{HP}$ vanishes and an extremal black hole dominates the grand canonical ensemble. To find our density of states, we extremize $\beta_{HP}(\Phi)(\Delta-\Phi Q)$ and find the bound for charged operators is
\begin{align}
\log \rho(\Delta,Q) \lesssim \frac{2\pi \Delta}{d-1}\sqrt{1-\hat{Q}^2/c^2}
\label{chargebound}
\end{align}
for $\Delta\geq|Q|/c$. The lower limit on the energies is the BPS bound for these black holes. Supersymmetry appears through considering Einstein-Maxwell as a consistent truncation of some supergravity theory. The fact that there cannot exist states lighter than the BPS bound $\Delta>Q/c $, can be seen from our bound eq. \eqref{chargebound}, which vanishes (more precisely, is $\mathcal{O}(1)$) in the BPS limit $\Delta= Q/c$. Unlike the previous subsection, the bound on the density of states \textit{at} the BPS limit vanishes. This is consistent with the nakedly singular nature of these BPS states.



\subsection{Single spin, uncharged operators}\label{spin}
For uncharged operators with a single spin, the dual bulk black hole is the $(d+1)$-dimensional Kerr black hole, analyzed first in \cite{Hawking:1998kw} for $d>2$. For $d=2$, we work with the spinning BTZ black hole \cite{BTZ}. The relevant thermodynamic parameters for these black holes are the uncharged single spin limit of sections \ref{d3bound}, \ref{AdS5}, \ref{AdS6}, \ref{AdS7} for $d=3,4,5,$ and $6$, respectively, where the relevant thermodynamical quantities are listed. 
The difference of regularized on-shell Euclidean actions becomes
\begin{align}
\Delta I_E=\frac{\beta_{d+1}\omega_{d-3}}{4G(d-2)\Xi}\left[r_+^{d-4}(r_+^2+a^2)(1-r_+^2)\right],
\end{align}
so black holes dominate for $r_+> 1$. The inverse temperature for the Hawking-Page transition is
\begin{align}
\beta_{HP}(\Omega) =\frac{2\pi}{d-2+\sqrt{1-\Omega^2}}.
\end{align}
To find the density of states, we extremize $\beta_{HP}(\Omega)(\Delta-\Omega J)$ with respect to $\beta_{HP}$ and find
\begin{align}
\label{KerrBound}
\log\rho(\Delta,J) \lesssim \f{2 \pi\Delta}{(d-3)(d-1)} \left(d-2-\sqrt{(d-3) (d-1) \hat{J}^2+1}\right),
\end{align}
where again $\hat{J}=J/\Delta$. The $d=3$ case is obtained by taking the limit. The unitarity bound is $\Delta\geq |J|$, which can also be understood as a BPS bound by taking the limit of zero $U(1)$ charge.

The result for $d=2$ agrees with the HKS bound \cite{Hartman:2014oaa}. It is notable that in this case, the bound from cosmic censorship agrees with the BPS bound, $\Delta - c/12 \geq |J|$ \cite{Hawking:1998kw}, where we have normalized $E_{\text{vac}} =-c/12$. However, the HKS bound allows states down to $\Delta = |J|$, which is the saturation point of the unitarity bound $\Delta  \geq |J|$. This only occurs in $d=2$: all higher-dimensional bounds obtained by our method will coincide with BPS bounds. 
Because of similarities with multiple spin operators derived in the next sections, we also note that the single spin bound may be written as
\begin{align}
\log\rho(\Delta, J) \lesssim \pi\Delta s
\end{align}
where $s$ is the smallest non-negative solution to 
\begin{align}
(d-2)s + \sqrt{s^2+4\hat{J}^2} =2.
\end{align}



\subsection{Multiple spin and zero charge operators}\label{spinspin}
\begin{figure}[t!]
\begin{center}
\includegraphics[scale=.43]{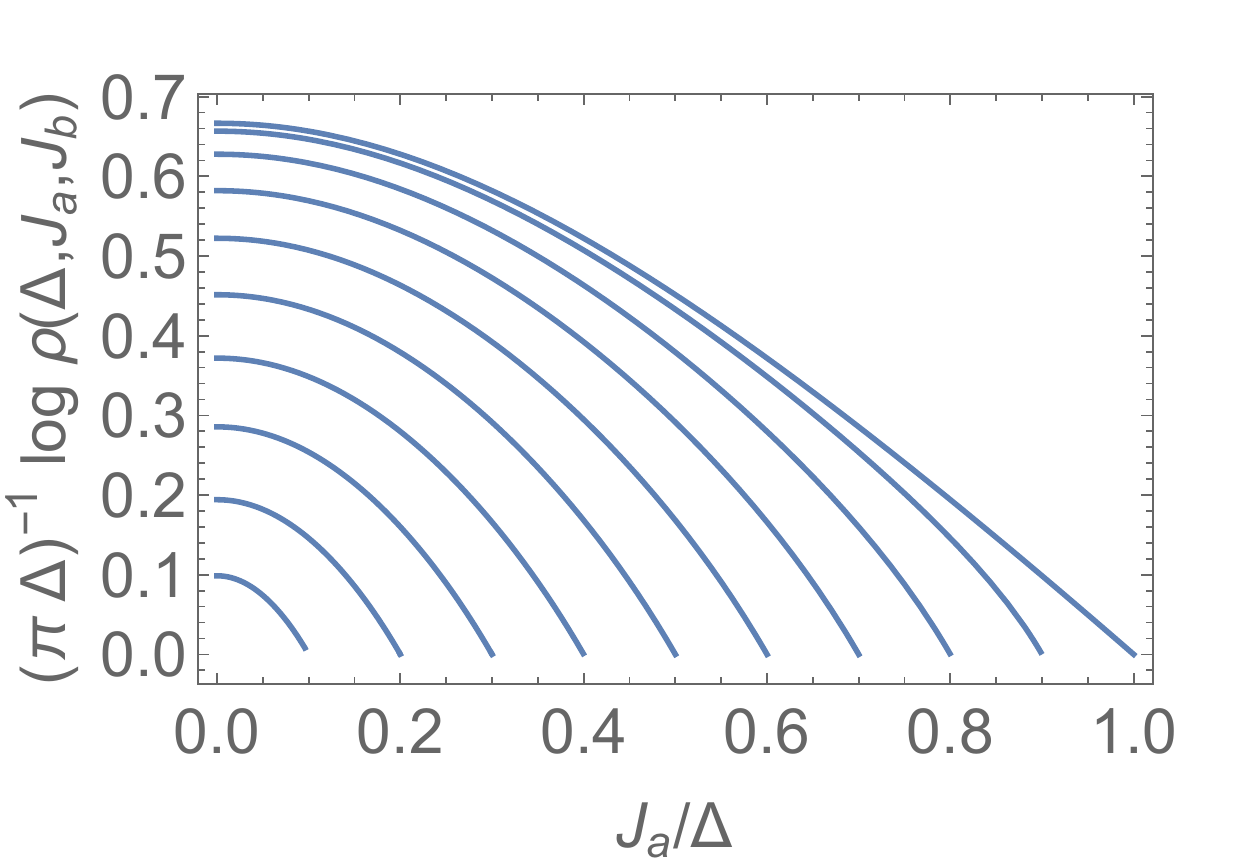}
\includegraphics[scale=.43]{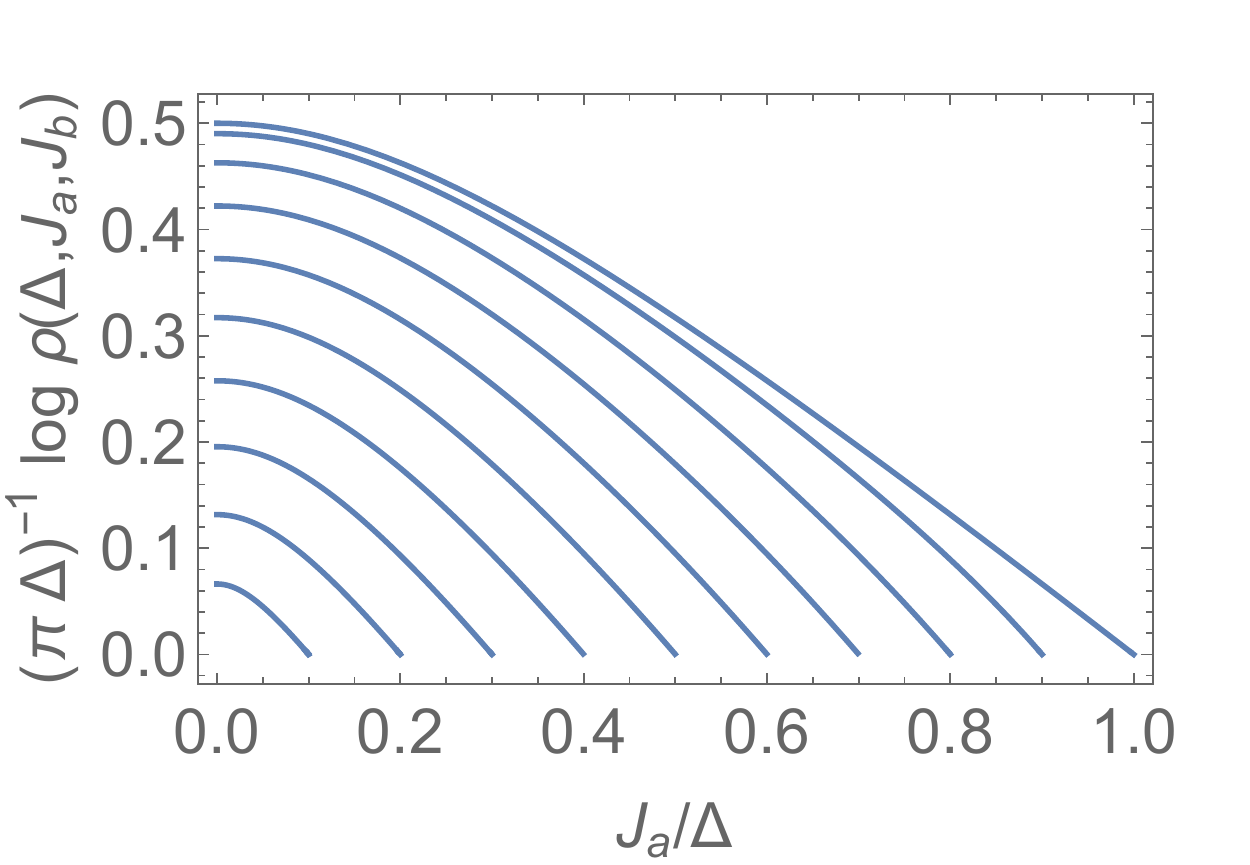}
\includegraphics[scale=.43]{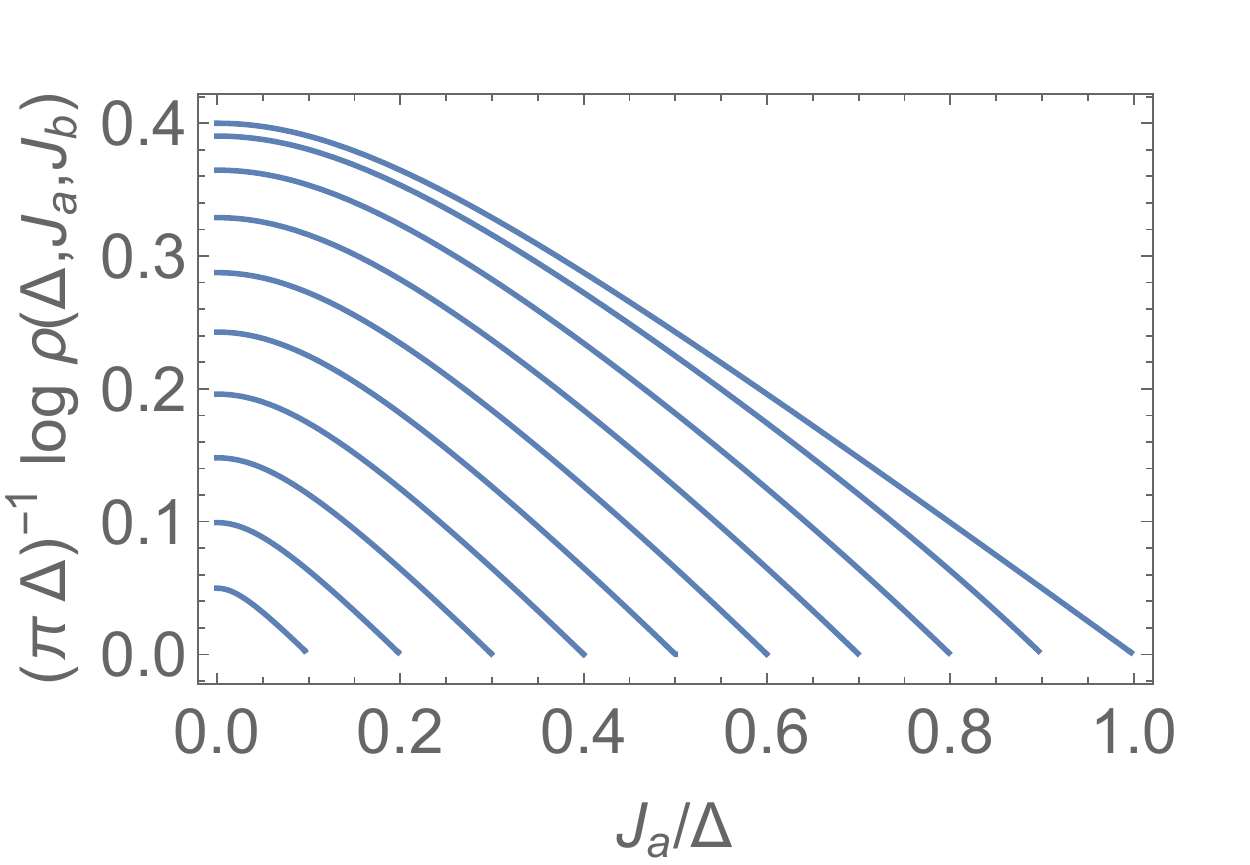}
\caption{The bound for operators with spins $J_a$ and $J_b$ in $d=4$ (left), $d=5$ (middle), $d=6$ (right). Curves correspond to $J_b/\Delta=0$ (rightmost) to $J_b/\Delta=.9$ (leftmost) in increments of .1}
\end{center}
\end{figure}
Analytic expressions are possible for two spins and zero $U(1)$ charge. Here, the bulk black holes are spinning AdS-Myers-Perry black holes in dimension $d>3$, whose metrics can be obtained from the zero charge limit of the gauged supergravity solutions\cite{Chong:2005hr,Chow:2008ip} in $d=4,5$ respectively and from the zero charge, two spin limit of \cite{Chow:2007ts} in $d=6$. The relevant thermodynamics as well as vacuum subtracted Euclidean actions are obtained in these limits from sections \ref{AdS5}, \ref{AdS6}, \ref{AdS7}.
Myers-Perry black holes dominate the grand canonical ensemble for $r_+>1$. The Hawking-Page temperature is
\begin{align}
\beta_{HP}(\Omega_a,\Omega_b) = \frac{2\pi}{(d-3)+\sqrt{1-\Omega_a^2}+\sqrt{1-\Omega_b^2}}.
\end{align}
We find that extremizing $\beta_{HP}(\Omega_a,\Omega_b)(\Delta-\Omega_aJ_a-\Omega_bJ_b)$ is equivalent to finding the smallest non-negative solution to
\begin{align}
\label{2spin}
(d-3)s + \sqrt{s^2+4\hat{J}_a^2}+\sqrt{s^2+4\hat{J}_b^2}=2
\end{align}
where $\hat{J}_i = J_i/\Delta$ and our bound is
\begin{align}
\log\rho(\Delta,J_a,J_b) \lesssim \pi \Delta s.
\end{align}

\noindent For completeness, we will solve eq. (\ref{2spin}) explicitly. First, define
\begin{align}
x = 1+\frac{(d-3)^2}{2}(\hat{J}_a^2+\hat{J}_b^2),\quad\quad y=\frac{3}{2}(\hat{J}_a^2-\hat{J}_b^2).
\end{align}

\noindent In $d=4$, the bound is
\begin{align}
\log\rho(\Delta,J_a,J_b) \lesssim \frac{2\pi\Delta}{3}\left(1-B_4+\sqrt{\frac{(B_4+2) ((2-B_4) B_4-6 x+8)}{B_4}}\right)
\end{align}
where
\begin{align}
A_4 &= \left(\sqrt{3 x^3 (3 x-4) y^2+6 ((x-6) x+6) y^4+y^6}+x^3+3 x y^2-6 y^2\right)^{1/3},\\
B_4 &= \sqrt{\frac{-2 A_4 x+A_4 (A_4+4)+x^2-1}{A_4}}\nonumber.
\end{align}

\noindent In $d=5$, the bound is
\begin{align}
\log&\rho(\Delta,J_a,J_b)\lesssim\\
&\frac{\pi \Delta}{6}\Biggl(6-x-\left(x^3-4 \left(\sqrt{36 y^4-3 x^3 y^2}+6 y^2\right)\right)^{1/3}-\frac{x^2}{\left(x^3-4 \left(\sqrt{36 y^4-3 x^3 y^2}+6 y^2\right)\right)^{1/3}}\Biggr).\nonumber
\end{align}

\noindent In $d=6$, the bound is
\begin{align}
\log\rho(\Delta,J_a,J_b) \lesssim \frac{2\pi \Delta}{15}\left(7-B_6-\sqrt{\frac{(2-B_6) \left(B_6^2+2 B_6-2(5x+4)\right)}{B_6}}\right),
\end{align}
where
\begin{align}
A_6&=\Biggl(9 (5 x+6) y^2-x^3+3 \sqrt{-3 x^3 (5 x+4) y^2+6 (5 x (5 x+18)+54) y^4-375 y^6}\Biggr)^{1/3},\\
B_6&=\sqrt{\frac{5 A^2 + 5 x^2 + 2 A (6 + 5 x) + 75 y^2}{3 A}}.\nonumber
\end{align}

\noindent One must be careful with these expressions to always take the principal root, which is generally complex, though the bound is always real for $|J_a|+|J_b|\leq\Delta$. For instance, in the no spin limit, $A_6\to \exp(i\pi/3)$ and $B_6\to 3$. Like in the previous section, there is a unitarity bound $|J_a| + |J_b| = \Delta$ which can be understood as a BPS bound by taking the limit of zero $U(1)$ charge. It can be shown $|J_a|+|J_b| \rightarrow \Delta$, only when $|J_i|,\Delta \rightarrow 0 $ or they both diverge. In the first case, our bound vanishes and is consistent with the bulk, while in the latter case the bound diverges and is saturated by the divergent entropy of the corresponding black hole. We close this section with a remark on the triply spinning case. Though it must be solved numerically, the bound on triply spinning operators can be obtained from the simple expression
\begin{align}
(d-4)s+\sqrt{s^2+4\hat{J}_a^2}+\sqrt{s^2+4\hat{J}_b^2}+\sqrt{s^2+4\hat{J}_c}=2.
\end{align}
The smallest non-negative solution to this expression gives our bound,
\begin{align}
\rho(\Delta,J_a,J_b,J_c) \lesssim \pi \Delta s.
\end{align}


\subsection{Single spin and single charge operators}\label{spincharge}
\begin{figure}[t!]
\begin{center}
\includegraphics[scale=.5]{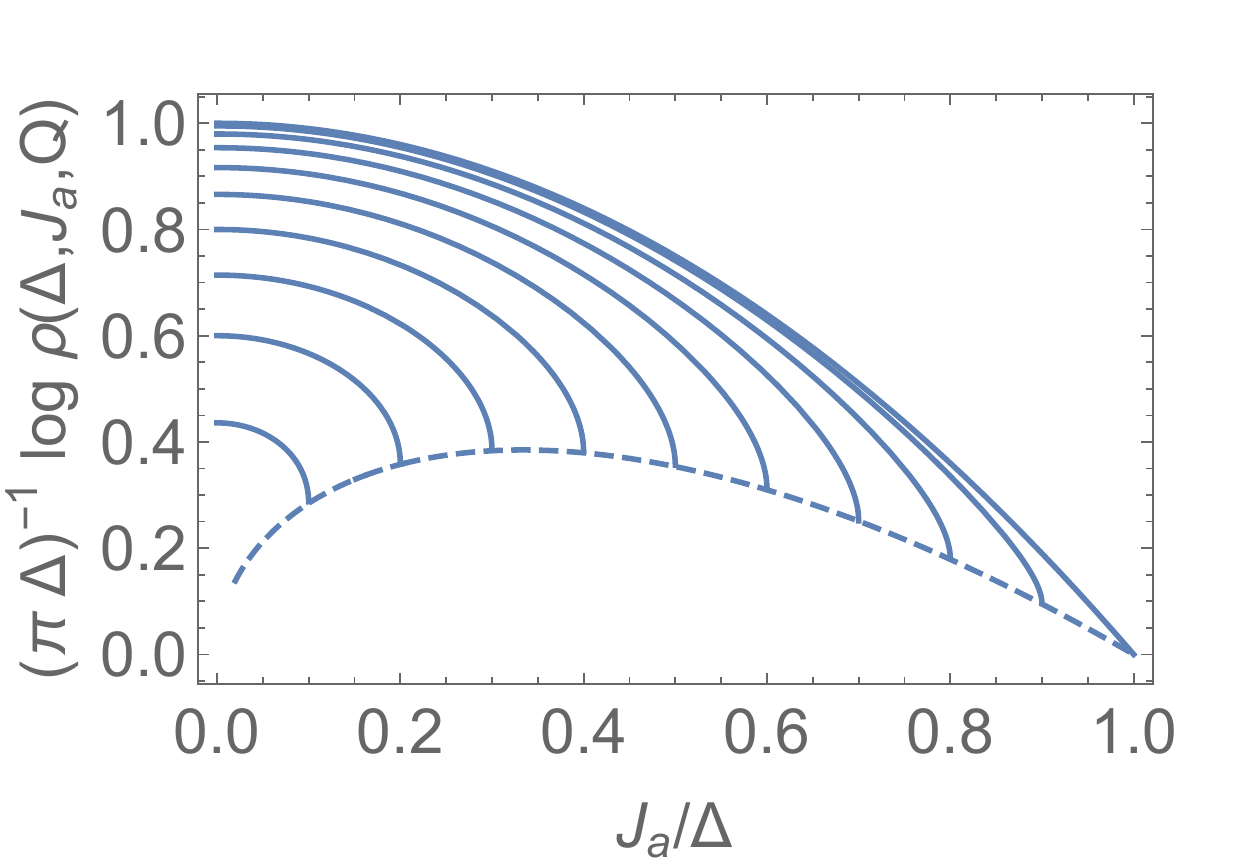}
\includegraphics[scale=.5]{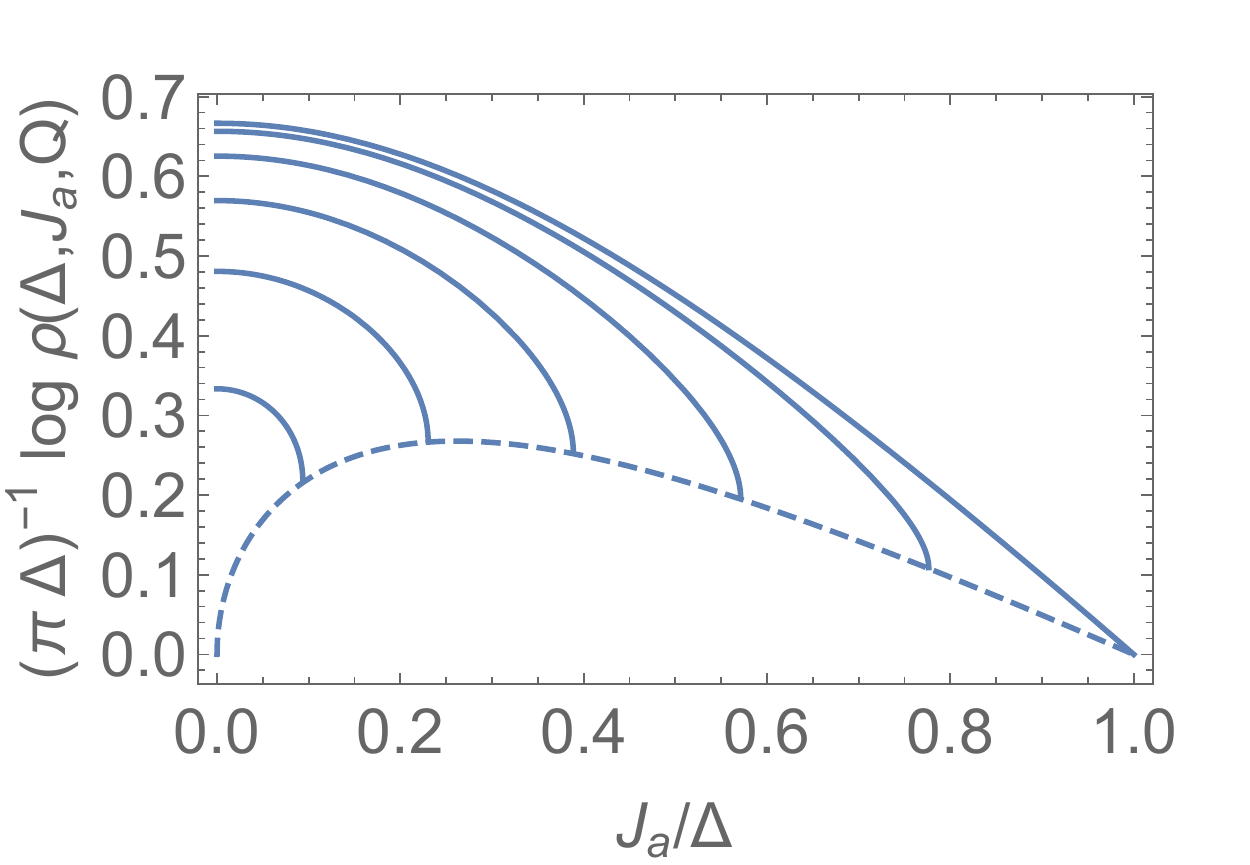}
\caption{\label{jq4.pdf}The bound for operators with spin $J_a$ and charge $Q$ in $d=3$ (left) and $d=4$ (right). From right to left, thick curves range from $Q/\Delta=0$ to $Q/\Delta=1$ (in $d=3$) or $Q/\Delta=1/\sqrt{3}$ (in $d=4$) in increments of $.1$. In both plots, the dashed line is the horizon entropy per mass of the BPS black hole $S_{BH}/\pi \Delta$.}
\end{center}
\end{figure}
\begin{figure}[t!]
\begin{center}
\includegraphics[scale=.5]{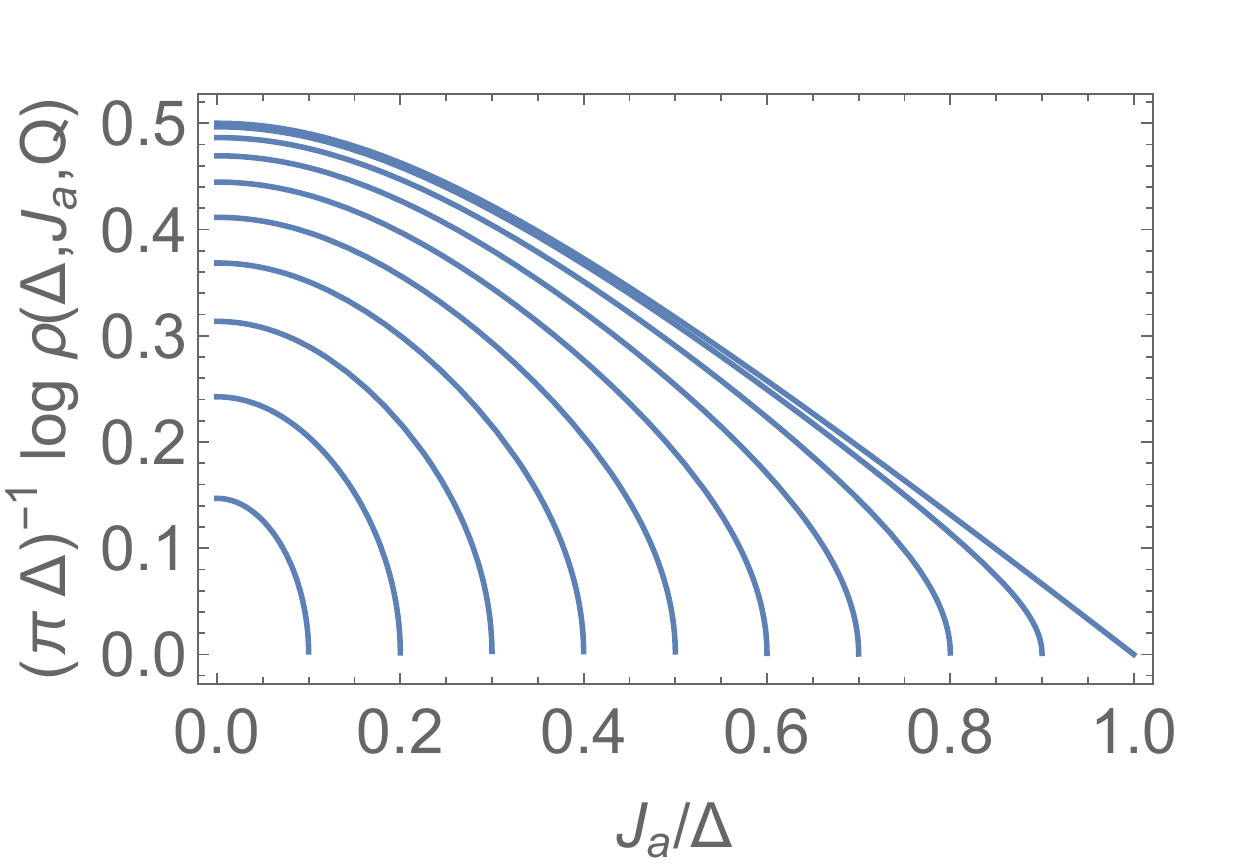}
\includegraphics[scale=.5]{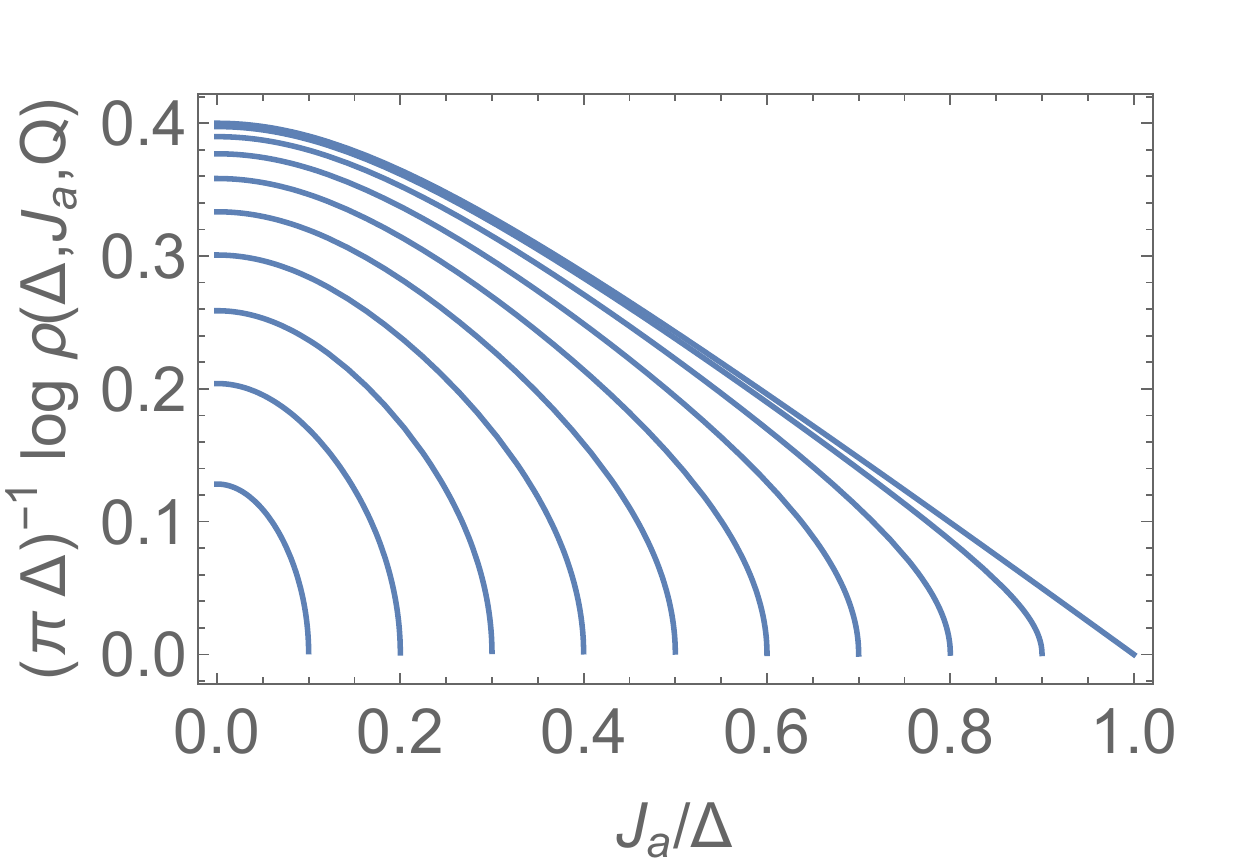}
\caption{\label{jq7.pdf}The bound for operators with spin $J_a$ and charge $Q$ in $d=5$ (left) and $d=6$ (right). From right to left, thick curves range from $Q/\Delta=0$ to $Q/\Delta=1$. The bound vanishes at the BPS limit $|J|+Q=\Delta$.}
\end{center}
\end{figure}

\noindent Bounds for single spin and single charge operators exist in $d>2$. We already derived the bound for $d=3$ in section \ref{d3bound}. In $d=4$, we take the single spin limit of the black hole in \cite{Chong:2005hr}. In $d=5$ and $d=6$, we choose the single spin and single charge black hole from \cite{Chow:2008ip} and \cite{Chow:2007ts}, respectively. It is worth noting that the generically spinning, charged black holes with AdS$_6$ and AdS$_7$ asymptotics are not pure Einstein-Maxwell, whose generically spinning solutions are not known in these dimensions, but are rather truncations of minimally gauged supergravity. Their zero-spin limit is not AdS-Reissner-Nordstr\"om and so this limit will not agree with section \ref{chargedanalytic}. Relevant thermodynamic quantities and vacuum subtracted Euclidean actions are listed in appendix B, in the single spin and single charge limit. As in $d=3$, it is easiest to find $\Phi(r_+,\Omega)$ at the Hawking-Page transition and then minimize over $\beta(r_+,\Omega)(\Delta - \Omega J - \Phi(r_+,\Omega)Q)$. In $d=4$, we have the odd feature (see section \ref{AdS5}) that $b=0$ does not imply $\Omega_b = 0$ or $J_b=0$. However, this choice gives a nice analytic bound which can be written purely in terms of $\Delta, J_a, Q$. Defining
\begin{align}
J_b=\Delta\frac{\left(\tilde{J}-1\right) \hat{Q}}{\sqrt{3} \hat{J}_a} \quad\text{and}\quad \tilde{J} = \sqrt{1+3\hat{J}_a^2}.
\end{align}
we have
\begin{align}
\rho(&\Delta,J_a,J_b,Q)\lesssim \frac{\pi  \Delta }{3 \left(1+\tilde{J}\right)^2} \\
&\times\left(\tilde{J} \left(1 -\tilde{J}\right)+\sqrt{\left(\tilde{J}(1-\tilde{J})+2+6 \hat{Q}^2\right)^2-12 \hat{Q}^2 \left(1+\tilde{J} \right)^2}+2+6 \hat{Q}^2\right)\nonumber\\
&\times\sqrt{\tilde{J} \left(2\tilde{J}+1\right)-6\hat{Q}^2-1+\sqrt{\left(\tilde{J} \left(1-\tilde{J}\right)+6\hat{Q}^2+2\right)^2-12\hat{Q}^2 \left(1+\tilde{J}\right)^2}}\nonumber
\end{align}
Notable in this bound is the BPS limit, $J_a+J_b + \sqrt{3}Q = \Delta$, which does not vanish but, as in $d=3$, reproduces the entropy of the corresponding BPS black hole,
\begin{align}
\text{max}\left[\rho\left(J_b+J_a+\sqrt{3} Q=\Delta\right)\right] = \frac{2 \pi  \Delta  \left(1-\hat{J}_a\right) \sqrt{\hat{J}_a \left(\tilde{J}-1\right)}}{\tilde{J}+3\hat{J}_a-1}.
\end{align}

\noindent In $d=5$, we get the bound
\begin{align}
\log \rho(\Delta, J, Q)&\lesssim \frac{\pi \Delta}{4}\frac{\left(9-\sqrt{72 \hat{J}^2-8 \hat{Q}^2+9}\right) }{9-9 \hat{J}^2+\hat{Q}^2}\\&\times \sqrt{9(1-\hat{J}^2)^2+\hat{Q}^2\left(\hat{Q}^2-10\hat{J}^2-8-2\sqrt{9+72 \hat{J}^2-8 \hat{Q}^2}\right)}.\nonumber
\end{align}
Here, the density vanishes in the BPS limit $|J|+Q=\Delta$. Finally, in $d=6$, our bound is
\begin{align}
\log \rho(\Delta, J, Q) &\lesssim\frac{2 \pi  \Delta}{15}\frac{\left(16-\sqrt{240 \hat{J}^2-15 \hat{Q}^2+16}\right)}{16-16 \hat{J}^2+\hat{Q}^2}\\&\times
 \sqrt{16 \left(1-\hat{J}^2\right)^2+\hat{Q}^2 \left(\hat{Q}^2-17\hat{J}^2-15-2\sqrt{16+240 \hat{J}^2-15 \hat{Q}^2}\right)}\,.\nonumber
\end{align}
Here too, the density vanishes in the BPS limit $|J|+Q=\Delta$. The vanishing at $|J|+Q=\Delta$ in $d=5,6$ is a consequence of the fact that BPS black holes only exist for $J_a,J_b,Q$ non-vanishing \cite{Chow:2007ts, Chow:2008ip}.

\subsection{Numerical Results}\label{numal}
\label{NumericalSection}

In the previous sections, we calculated analytic bounds for operators with up to three parameters. To obtain bounds for operators with four or more parameters, we must resort to numerics. With $\left \lfloor d/2 \right\rfloor$ angular potentials and one chemical potential, the Hawking-Page temperature is a $\left\lfloor d/2\right\rfloor+1$ dimensional hypersurface. For a given set of charges, $\{J_1, J_2,...,J_{\left\lfloor d/2\right\rfloor},Q\}$, we then numerically find the minimum value of 
\begin{align}
\beta_{HP}(\Omega_1,\Omega_2,...,\Omega_{\left\lfloor d/2\right\rfloor},\Phi)\left[1-\sum_{i=1}^{\left\lfloor d/2\right\rfloor}\hat{J}_i\Omega_i -\hat{Q}\Phi\right]
\label{minimization}
\end{align}
where for simplicity we have scaled out an overall factor of $\Delta$ so that all charges fall in a finite range. In the full ensemble, the BPS bound is $\Delta = \frac{1}{c}|Q|+\sum_{i=1}^{\left\lfloor d/2\right\rfloor}|J_i|$.\footnote{For the $d=5,6$ supergravity solutions, the normalization of the charge is such that $c=1$ rather than the $c$ defined for the Reissner-Nordstr\"om black holes} For energies below this bound, the density of states vanishes at leading order in $N$. 

\begin{figure}[H]
\begin{center}
\includegraphics[scale=.39]{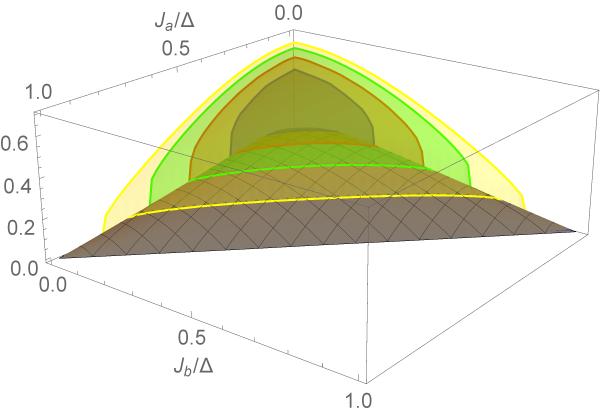}
\hspace{12mm}
\includegraphics[scale=.39]{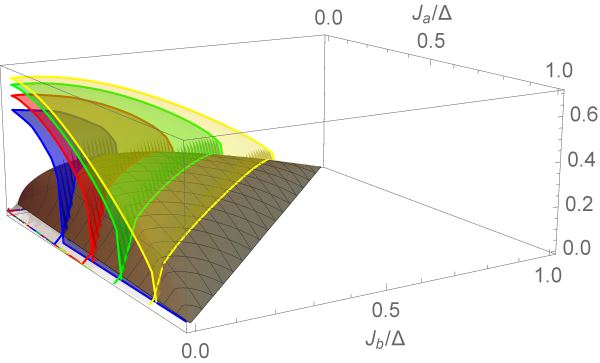}
\caption{\label{4dspincharge}\textbf{$d=4$}: $Q/\Delta=.4, .3, .2, .1$ (blue, red, green, yellow). The BPS condition is $\Delta = J_a+J_b+\sqrt{3}Q$.}
\includegraphics[scale=.39]{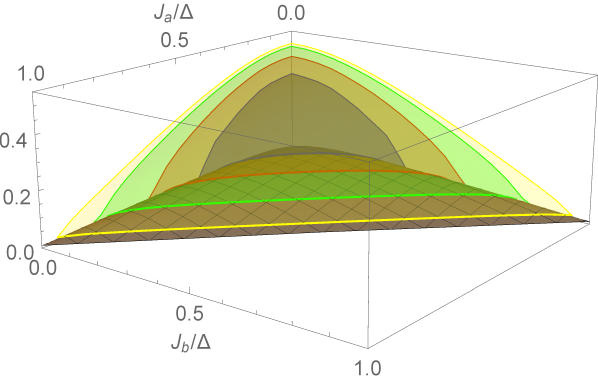}
\hspace{12mm}
\includegraphics[scale=.39]{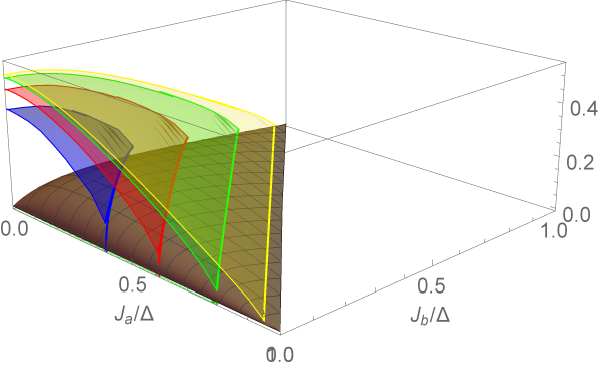}
\caption{\textbf{$d=5$}: $Q/\Delta=.6, .4, .2, .1$ (blue, red, green, yellow). The BPS condition is $\Delta = J_a+J_b+Q$.}
\includegraphics[scale=.39]{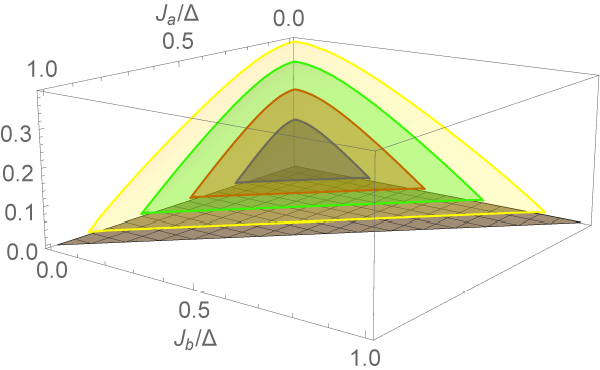}
\hspace{12mm}
\includegraphics[scale=.39]{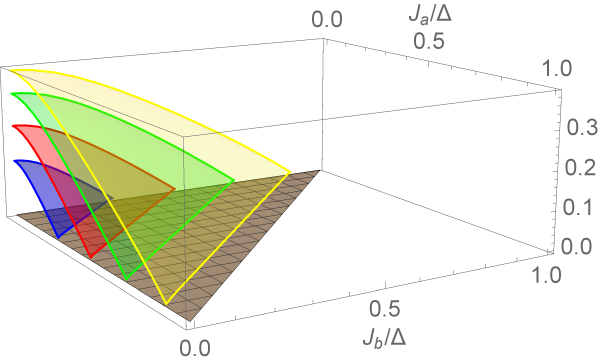}
\caption{$d=6:$ $Q=0$ and $J_c/\Delta=.7, .5, .3, .1$ (blue, red, green, yelow). There are no BPS black holes with vanishing $U(1)$ charge, so the gray surface is $S_{BH}=0$, where our bound implies $O(1)$ degeneracy of states.}
\includegraphics[scale=.39]{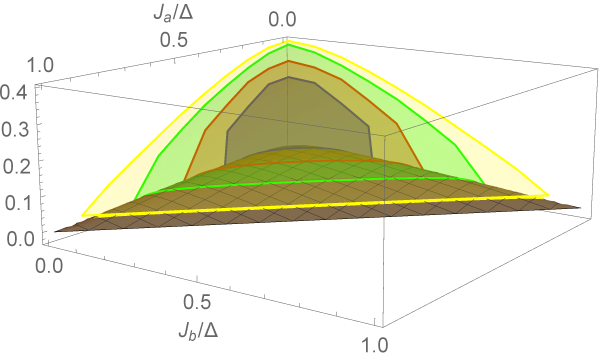}
\hspace{12mm}
\includegraphics[scale=.39]{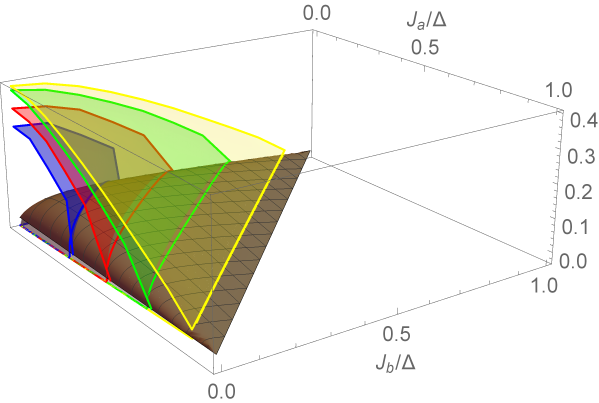}
\caption{\label{6dspincharge}$d=6$: $J_c=0$ and $Q_c/\Delta=.7,.5,.3, .1$ (blue, red, green, yelow). The BPS condition is $\Delta = J_a+J_b+Q$. $\\$}
\emph{Plots of $(\pi\Delta)^{-1}\log \rho[\Delta,J_a,J_b,(J_c),Q]$. Gray surfaces are $S_{BH}/(\pi\Delta)$ for corresponding BPS black holes which coincide with our bound at the BPS condition. Beyond this surface, the bound vanishes and no states are allowed (color online).}
\end{center}
\end{figure}

Because the equations we need to solve are algebraic, no sophisticated numerical techniques are necessary. We discretize the thermodynamic potentials and (hatted) parameters which have finite range. Angular potentials are bounded from above by the speed of light of the boundary ESU, $\Omega_i = 1$ and the electric potential is bounded from above by cosmic censorship. The spins and electric charges, scaled by the energy, also have finite range, typically $\{\hat{J}_i,\hat{Q}\} \in [0,1]$ but this depends on the normalization of $A_\mu$. The exact limits can be found in the appendix using the BPS bounds. We divide these intervals into equally spaced grids of $N=100$ points. For each grid point labeled by the potentials' $(\left\lfloor d/2\right\rfloor+1)$ coordinates, we used the built-in ``NSolve" function in Mathematica to obtain the black hole radius at the Hawking-Page transition giving us the critical surface defined in section \ref{methodsec}. Once obtained, we calculate eq. (\ref{minimization}) for each grid point in the spins' and charge's $(\left\lfloor d/2 \right\rfloor+1)$ coordinates. Then, for each point $\{\hat{J}_i, \hat{Q}\}$ we searched for the minimum value of eq. (\ref{minimization}) over the potentials, imposing the lower bound of zero. Because eq. (\ref{minimization}) is exponentiated for the density of states, the lower bound determines where a single state is allowed--this is the BPS/unitarity bound of the CFT. Beyond this point (or curve), our procedure allows no states.

 As checks on the numerics, we verified that our curves did not vary appreciably as a function of the grid sizes and that they agreed with the analytic results in the previous subsections. In figures \ref{4dspincharge} through \ref{6dspincharge}, we plot the bound on the density of states in each dimension. Notable in these plots is the entropy of BPS black holes, plotted as a gray surface. Our bounds end on this surface, giving the entropy of these black holes, and then vanish, marking the BPS bound of the CFT. Furthermore, as we pointed out in section \ref{spincharge}, with only one spin and charge, there are no BPS black holes and hence the gray entropy surface vanishes.



\section{BPS, cosmic censorship, and sparseness bounds}\label{bpssec}
\begin{figure}[t!]
\begin{center}
\includegraphics[scale=.37]{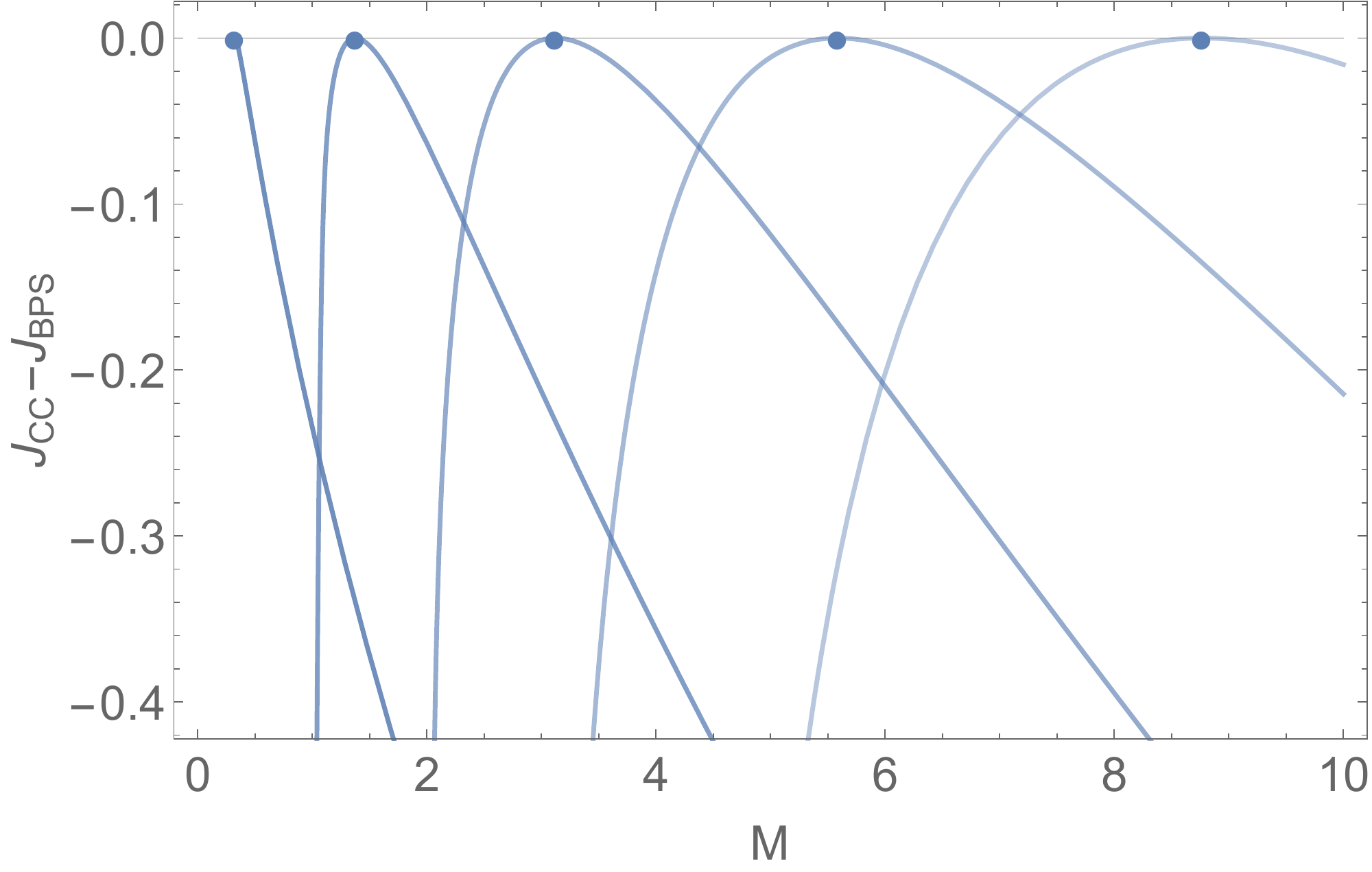}
\hspace{.3cm}
\includegraphics[scale=.37]{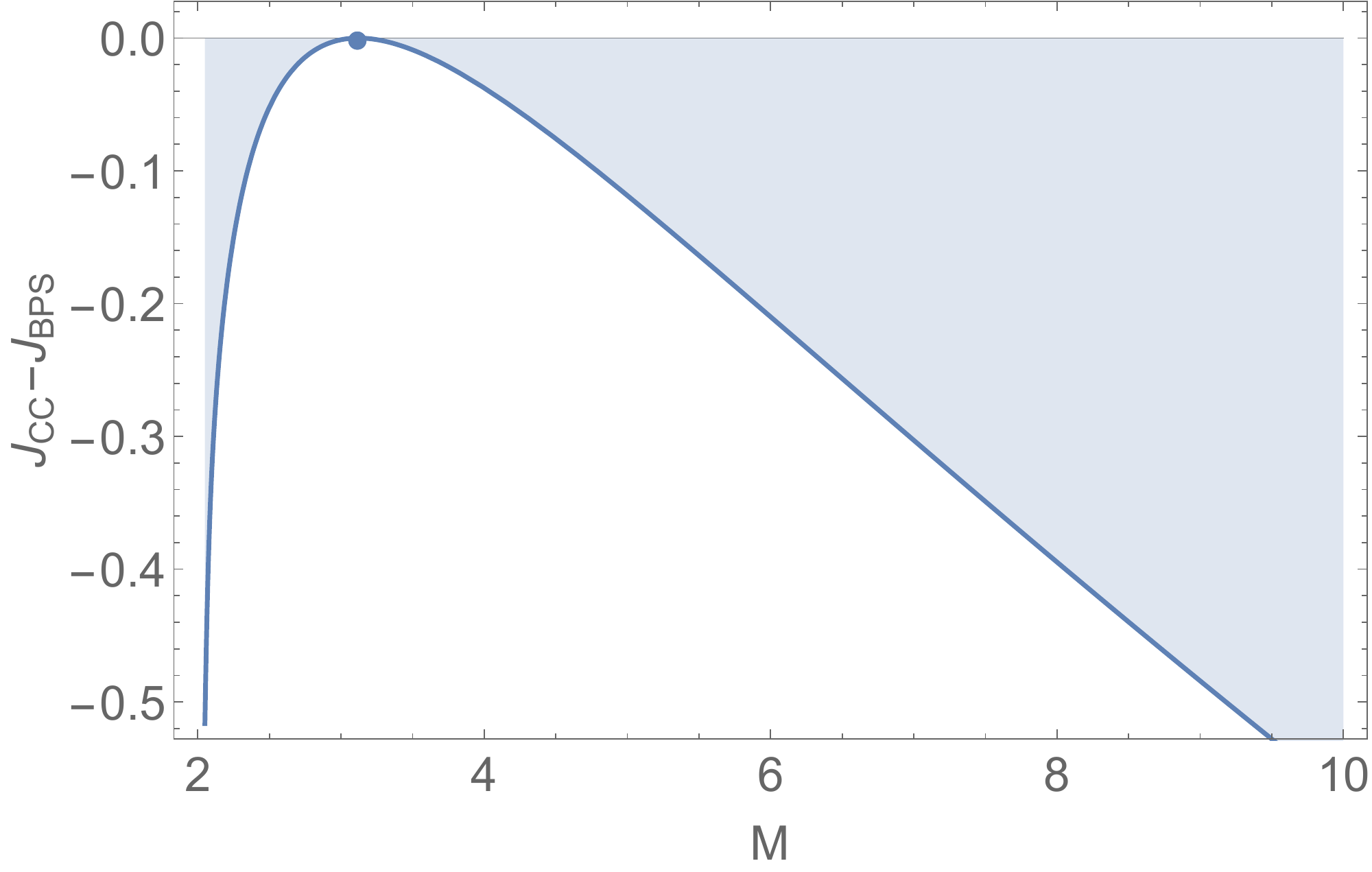}
\caption{\label{4Dccbound} (Left) The difference between the BPS and cosmic censorship bound for four-dimensional Einstein-Maxwell-AdS. Curves correspond to fixed $Q=.3, .9, 1.5, 2.1, 2.7$ (left to right, darker to lighter). Circles indicate locations where extremal black holes satisfy the BPS condition $M=|J|+Q$ and BPS states are otherwise superextremal. (Right) The $Q=1.5$ curve. Shaded region corresponds to CFT states with superextremal bulk duals that are not excluded by our bound. The feature $J_{BPS} > J_{CC}$ except at isolated points where $J_{BPS}=J_{CC}$ is characteristic of charged, spinning solutions in $d=3, 4$ with at least one spin and $d=5, 6$ with at least two spins. Without both charge and spin the inequality is never saturated. }
\end{center}
\end{figure}

In previous sections, we saw that our bound on the density of states vanishes at leading order in $N$ for states that violate the BPS  bound in $d>2$. This is intriguing since we generically considered non-supersymmetric (Einstein-Maxwell) theories, without using any embedding into supergravity. The appearance of a coarse BPS condition suggests that bulk thermodynamics knows about the consistent supergravity extension. Its appearance is due to the upper bounds on the chemical potentials in the confined phase of strongly coupled holographic theories. To see this, consider the case of finite temperature and a single angular potential. The confined phase always satisfies $\Omega \leq 1$, which means minimizing $\exp\left(\beta(E-\Omega J)\right)$ in the confined phase will give zero for $J>E$, since then we can pick $\Omega = 1$ and $\beta \rightarrow \infty$. Had the confined phase admitted $\Omega > 1$, then our bound would rule out states with $J>J_c$ where $J_c < E$. 

The bulk gravitational theory also has an additional bound -- the cosmic censorship (CC) bound, that arises by demanding that there are no naked singularities. In general these two bounds are different: for $\Delta_{BPS}$ the lower bound implied by the BPS bound and $\Delta_{CC}$ the lower bound implied by cosmic censorship, we have $\Delta_{BPS} < \Delta_{CC}$ for fixed $U(1)$ charge or fixed spins, i.e. BPS states violate cosmic censorship. In the case with both $U(1)$ charge and spin, there is a line $J(Q)$ along which $\Delta_{BPS} = \Delta_{CC}$ if there is at least one spin in $d=3, 4$ and at least two spins in $d=5, 6$ (see figure \ref{4Dccbound}). We find that in the cases where $\Delta_{BPS}$ is strictly smaller than $\Delta_{CC}$, our bound vanishes at the BPS bound, while in the case where the BPS bound coincides with the CC bound, the maximum of our bound reduces to the entropy of the extremal black hole. Masses between cosmic censorship and BPS must have superextremal bulks but our bound allows for an $O(N^k)$ number of states in this range. In this section, we quickly review the BPS and CC bounds to compare to the sparseness bounds we obtain from the Hawking-Page transition.

For singly spinning black holes, as mentioned in previous sections, there is a unitarity bound that can also be understood as a $Q\rightarrow 0$ limit of a BPS bound,
\begin{equation}
\Delta  \geq J\,.
\end{equation}
Thus $\Delta_{BPS}=J$ becomes the lower bound on the allowed energy levels. This energy is also found to be strictly less than the cosmic censorship bound. In the limit $\Delta\rightarrow\Delta_{BPS}$, we find that our bound gives vanishing degeneracy of states at leading order in $N$. This is consistent with the fact that the only uncharged spinning BPS states are superextremal, and hence have $O(1)$ entropy. There are no extremal black holes with only one spin in $d \geq 5$, which is easily seen from the emblackening factor of the Kerr metric in Boyer-Lindquist coordinates \cite{Hawking:1998kw}
\begin{equation}
\Delta_r = (r^2+a^2)(1+r^2)-2mr^{4-d}.
\end{equation}
However, there is still a ``speed limit," $a\to 1$, required for stable bulk black holes.

For singly charged black holes, the BPS bound is given by
\begin{equation}
\Delta \geq \sqrt{\frac{d-1}{2(d-2)}}Q.
\end{equation}
At fixed charge, this energy is strictly less than the CC bound. Again we find that as $\Delta\rightarrow\Delta_{BPS}$ our bound gives vanishing degeneracy of states at leading order in $N$. 
For non-spinning black holes in Einstein-Maxwell theory, the BPS bound is only rigorously known in $d=3,4$ where embeddings into supergravity theories have been found. 
 
The same qualitative results are true for charged spinning black holes -- at fixed charges the BPS energy is less than or equal to the cosmic censorship bound on energy. In the case of single charge in $AdS_{d+1}$, single spin single charge or double spin in $AdS_6$ and $AdS_7$, BPS states are always superextremal, and we find that our bound vanishes in the BPS limit. However superextremal states that lie between the BPS bound and the cosmic censorship bound, are nakedly singular and have $O(1)$ entropy, but our bound allows for $O(N^k)$ states.

In the case of single spin single charge in $AdS_4$, single or double spin single charge in $AdS_5$, double spin single charge in $AdS_6$, and double spin or triple spin single charge $AdS_7$, an extremal black hole that saturates the BPS bound exists for specific values of $\{\Delta,J_i,Q\}$. In such cases, the maxima of our bound reduces to the entropy of the extremal black hole. For generic values of $\{\Delta,J_i,Q\}$ between the BPS bound and the CC bound, the black hole is superextremal and has $O(1)$ entropy, but our bound still allows for $O(N^k)$ states. These features are shown for the four-dimensional Einstein-Maxwell-AdS theory in figure \ref{4Dccbound}. For fixed $Q$ and $\Delta$, it is clear that there exist states with $J_{CC}<J<J_{BPS}$ which are allowed by our bound but must be superextremal. That such states can be allowed is not surprising considering the stability of AdS black holes. Charged rotating black holes can often be obtained via dimensional reduction of spinning supergravity black holes in higher dimensions (see sections \ref{AdS6} and \ref{AdS7}). Spinning black holes have superradiant instabilities by which the black hole should decay to the most stable spinning charged states (i.e. BPS). This instability is reflected in the lower dimension because the extremal black hole is not supersymmetric and hence unstable. Our bound allows for a finite number of superextremal states to which the extremal black hole can decay. Recent work relating BPS and cosmic censorship bounds can be found in \cite{Horowitz:2016ezu, Crisford:2017gsb}.



\section{Comments on the high-lying spectrum}\label{cvsec}
Our bounds imply a range of vacuum dominance ($\beta > \beta_{HP}$) that matches the phase structure of Einstein gravity in the bulk. It is interesting to ask if the high-temperature phase structure ($\beta<\beta_{HP}$) can be reproduced without additional assumptions. This is what was done in \cite{Hartman:2014oaa, Belin:2016yll} by using modular invariance of the torus partition function. Since we are considering theories on $S^{d-1}$, where ordinary modular invariance is absent, we need other tools. 

We begin with an analysis of the Cardy-Verlinde formula \cite{Verlinde:2000wg}, which was proposed as a higher-dimensional analog to the Cardy formula on $S^{d-1}$:
\begin{equation}\label{cvform}
S=\frac{2\pi R}{d-1}\sqrt{E_{cas}(2E-E_{cas})}=\frac{4\pi R}{d-1}\sqrt{E_{subext}E_{ext}}.
\end{equation}
$R$ is the radius of $S^{d-1}$ of the CFT, $E=E_{ext} + E_{subext}$, and $E_{cas} \equiv 2E_{subext}$. $E_{ext}$ and $E_{subext}$ are the extensive and subextensive pieces of the thermodynamic energy. This formula reproduces the entropy of AdS-Schwarzschild black holes above the Hawking-Page transition but is known to fail for generic theories \cite{Kutasov:2000}.

A very important aspect of this formula is that, unlike the ordinary Cardy formula, it is canonical in nature. $E_{cas}$ is in no sense the ground state energy of the theory---as stated above, it is calculated by extracting the subextensive piece from $E\equiv\langle E \rangle$. That $E_{cas}$ cannot be a single energy level is clear by matching to the high-energy scaling $S \sim E^{(d-1)/d}$, which shows that $E_{cas}\sim E^{(d-2)/d}$ at leading order; in particular it has to scale with $E$. Furthermore, to compute $E_{cas}$, one has to have knowledge of $\log Z$ since $\langle E \rangle = -\partial_\beta \log Z$. But this means one already has knowledge of $S = (1-\beta \partial_\beta) \log Z$. So, the Cardy-Verlinde formula should be understood as a repackaging of thermodynamic quantities into a suggestive form. If not for the similarity to the ordinary Cardy formula it would be essentially meaningless. The parameters appearing in the Cardy formula, on the other hand, do not require knowledge of the thermodynamics. 
To have the Cardy-Verlinde formula reduce to the Cardy formula for $d=2$, as is often stated, one has to shift the definition of $E_{cas}$ by the vacuum energy $-c/12$. 

Nevertheless, the fact that the thermodynamic quantities can be repackaged in this way for holographic theories is nontrivial. It is then natural to ask how general it is---does it depend on the field theory manifold? Can one add potentials for electric charge and angular momentum? It turns out the formula fails for a holographic theory on flat slices, like a torus. This is because $E_{cas} = 0$ for such theories, making the formula meaningless. A constant shift only works for $d=2$, since for $d>2$ it would give incorrect asymptotic scaling $S \sim \sqrt{E}$. For this case, one has to instead use the higher-dimensional Cardy formula, which can be derived from modular invariance and is true for generic conformal theories \cite{Shaghoulian:2015kta, Shaghoulian:2015lcn}. On hyperbolic slices, it was shown that the formula fails but can be fixed by defining $E_{sub} = \frac{E_{cas}}{2k}$ \cite{Cai:1},
where spherical slices have $k=+1$ and hyperbolic have $k=-1$. With this definition, $E_{sub}$ is strictly positive. While no explanation was given for this substitution, we will use a high-temperature effective field theory to explain this result at the end of this section.

The formula fails generally when chemical potentials are added, although it can be fixed by making appropriate modifications in some cases. It has been shown that the entropy of Reissner-Nordstr\"om is reproduced by the Cardy-Verlinde formula on substituting $E_{ext}$ for $E_{ext} - \frac{\Phi Q}{2}$, where $\Phi$ and $Q$ are the $U(1)$ potential and the electric charge respectively\cite{Cai:Q}. While for Kerr-Newman black holes, thermodynamic quantities defined with respect to an asymptotically rotating frame can be shown to satisfy the Cardy-Verlinde formula\cite{Cai:2}. However in these modified definitions, $E_{ext}$ loses its meaning as being the extensive part of the energy. For more complicated solutions like multi-charged or multiply-spinning black holes in gauged supergravity models, one can still fix the Cardy-Verlinde formula by making changes to $E_{cas}$ and $E_{ext}$\cite{Cai:1}, however these changes are quite complicated in terms of the CFT thermodynamic quantities \cite{Gibbons:2005vp}. Thus, there does not seem to exist a universal modification that works for every case. While it is tempting to think the form of the Cardy-Verlinde formula implies a connection between high-lying and low-lying states, the difficulties outlined above, coupled with the fact that $E_{cas}$ is not a fixed low-lying energy, suggest otherwise.

Two approaches, which we will point out but not pursue, are to investigate the notions of ``emergent circles" \cite{Shaghoulian:2016gol} and ``detachable circles" \cite{Horowitz:2017ifu}. In this context, the notion of emergent circles says:
\be
Z\left[S^1_{2\pi/k\, \rightarrow\, 0} \times S^{2n+1}/\mathbb{Z}_{p\,\rightarrow\, \infty}\right] = Z\left[S^1_{2\pi/p\,\rightarrow \,0} \times S^{2n+1}/\mathbb{Z}_{k\,\rightarrow\, \infty}\right],\qquad p/k \text{ fixed }
\ee
The quotient is performed on the Hopf fiber for the odd-dimensional sphere represented as a circle fibered over $CP^n$. In this highly lensed limit, there is an emergent modular invariance that appears, since a highly lensed sphere behaves like $S^1 \times CP^n$ for the purpose of leading-order thermodynamics. Coupling this with the special pattern of center symmetry breaking of strongly coupled holographic CFTs \cite{Shaghoulian:2016xbx} may give an avenue to relating the theory on $S^{2n-1}/\mathbb{Z}_{p\rightarrow \infty}$ back to the theory on $S^{2n-1}$. For $n=3$ there is even a nontrivial Hawking-Page phase structure in the bulk with calculable $\beta(p)_{HP}$ that can be used to provide a bound on the density of states $\rho(E)\lesssim e^{\beta(p)_{HP}E}$ on $S^3/\mathbb{Z}_p$, connecting the round sphere $p=1$ to the case with an emergent modular invariance $p\rightarrow \infty$. 

The notion of ``detachable circles" in this context relates a finite-temperature conformal theory on $S^{d-1}$ to the theory on $\mathbb{H}^{d-1}/\mathbb{Z}$ at some inversely related temperature:
\be
ds^2 = d\chi^2 + d\theta^2 + \sin^2 \theta d\Omega_{d-3}^2 + \cos^2 \theta d\phi^2 \rightarrow \f{d\chi^2 + d\theta^2 + \sin^2 \theta d\Omega_{d-3}^2}{\cos^2\theta} + d\phi^2\,.
\ee
By restricting our theory to be gapped at finite temperature (which is the generic situation), we can use the effective theory approach introduced in \cite{meup}. This approach allows us to write down the following effective action for the theory dimensionally reduced over the thermal circle:
\be
\log Z(\beta) = \int d^{d-1} \sqrt{h} \left(c_0\f{1}{\beta^{d-1}}+c_1\f{R^{(1)}}{\beta^{d-3}}+c_2 \f{R^{(2)}}{\beta^{d-5}}+\dots\right)
\ee
This is to be understood as a perturbative expansion around $\beta \rightarrow 0$. Powers $R^{(n)}$ are to be understood as all possible combinations of contractions of the Riemann tensor, with e.g. different coefficients between $R_{\mu\nu}R^{\mu\nu}$ and $R^2$ which are suppressed for simplicity. 

This effective theory makes clear that the high-temperature theory on a hyperboloid is related to the high-temperature theory on the sphere by sign flips in the terms of the effective theory with odd powers of curvatures. Certain large-$N$ theories may have a sufficiently extended range of validity for this effective theory such that we can relate the theory on $\mathcal{H}^{d-1}/\mathbb{Z}$ back to the theory on $S^{d-1}$. This effective theory also explains why the Cardy-Verlinde formula works for hyperbolic slices with the definition $E_{sub}=\f{E_{cas}}{2k}$: this is a simple way to achieve the sign flips implied by the effective theory. 

\section{Conclusion}
In this paper, we derived quantitative sparseness conditions on holographic CFTs with a semiclassical Einstein dual. To arrive at these conditions, we used the fact that there generically exists a Hawking-Page transition between vacuum AdS and a large asymptotically AdS black hole at a particular temperature and set of thermodynamic potentials. Such a phase transition implies a discontinuous jump in the free energy from $O(1)$ to $O(N^k)$ and hence the CFT can only support a finite number of states before it deconfines. The difficulty in satisfying such bounds comes from the fact that interactions tend to sparsify a spectrum, so generic weakly interacting theories have dense spectra which violate our bounds. 

An interesting aspect of these bounds is that that $\log\rho = O(1)$ for masses below the BPS bound. In situations where a bulk BPS black hole exists at the bound, its entropy saturates our bound, which then discontinuously drops to $O(1)$ consistent with the bulk. It is interesting to see the appearance of the BPS bound in the cases with $U(1)$ charge without inputting supersymmetry. 

Sparseness assumptions figure prominently into simplifying limits of conformal bootstrap techniques. The usual style of argument is that a sufficiently sparse spectrum allows you to pick up only the contribution of the vacuum in a particular OPE expansion. This was most recently utilized in the bootstrap approach \cite{Jafferis:2017zna} to the ``large charge" expansion \cite{Hellerman:2015nra, Monin:2016jmo}. It would be interesting to explore the connection of our quantitative sparseness bounds to these bootstrap techniques. 


A sparse low-lying density of local operators is often invoked as a requirement for a CFT to have a semiclassical Einstein dual, but for $d>2$, a precise definition of ``sparseness" was lacking. In this work we have provided a quantitative sparseness bound on the allowed density of local operators in the CFT. This bound enforces vacuum dominance of the gravitational path integral at low temperatures. It is a sharp diagnostic for how much interactions have to ``sparsify" a spectrum, since it is \emph{violated} by weakly coupled holographic theories. It would be interesting to connect this sparseness condition to a \emph{different} sparseness condition, the gap to the higher-spin operators  \cite{Heemskerk:2009pn}, both of which need to be satisfied for a weakly coupled Einstein gravity dual, and both of which are violated for weakly interacting holographic CFTs in $d>2$.  



\vskip 1cm
\centerline{\bf Acknowledgements}
\vskip .5cm 
It is a pleasure to thank Alexandre Belin, Aleksey Cherman, and Thomas Hartman for discussions. This work was supported in part by NSF grant no. PHY-1316748 and Simons Foundation Grant 488643. The work of ES was performed in part at the Aspen Center for Physics, which is supported by NSF grant no. PHY-1607611. EM was supported in part by NSF grant no. PHY-1504541. MS was supported in part by the NSF grant PHY-1316699 

\appendix
\section{Black hole entropy from deconfining phase transitions}\label{bhentropy}
As we saw throughout this paper, our bounds on the density of states are saturated by the black hole at the deconfining phase transition. We can invert this logic to produce a derivation of black hole entropy from field-theoretic considerations. Since we would have to input a deconfinement temperature (and more assumptions) in the general case, let us focus on $d=2$ where we can get by with minimal assumptions. 

Assume a large-$c$ CFT in $d=2$ has a single first-order deconfining phase transition. By modular invariance it must occur at $\beta = 2\pi$. We use a normalization consistent with modular invariance,  $E_{\text{vac}}=-c/12$. We also know from the modular bootstrap that $\langle E \rangle_{\beta = 2\pi} = 0$ \cite{Hellerman:2009bu}.  By the generic description of first-order phase transitions as an exchange of saddles, we can use the vacuum energy and $\langle E \rangle_{\beta=2\pi} = 0$ to deduce that $\langle E \rangle_{\beta=2\pi-\epsilon} = c/12$ up to corrections in $\epsilon$. Since $\Delta_{c} - S_{c}/\beta_{c} = O(1)\implies S_{c} = \beta_{c} \Delta_{c}=\beta_{c}(E_{c}+c/12)$ at leading order in $c$, this gives us a prediction for the thermal entropy, where we have deduced $E_{c}=c/12$ and $\beta_{c}=2\pi$ purely from field-theoretic considerations. Notice that ``$c$" is doing double duty here. For $\beta = 2\pi-\epsilon$ we are in the deconfined phase of a large-$c$ theory, so we can coarse grain to translate into a density of states as in \cite{Hartman:2014oaa}. Altogether we have the formula 
\be
\log \rho(E = c/12) = \pi c/3\,.
\ee
This agrees precisely with the bulk, where the ensemble is dominated by a BTZ black hole below $\beta = 2\pi$, and so by continuity the density of states at $\beta = 2\pi$ is given by the Bekenstein-Hawking entropy of the BTZ black hole at the Hawking-Page transition. One could also dispense of the assumption that the transition is first order and so described by an exchange of saddles to obtain a formula of the sort $\log \rho(E) = 2\pi E$ where $E\equiv \langle E \rangle_{2\pi-\epsilon}$. 

Notice that the Cardy formula $\log\rho(E) = 2\pi \sqrt{cE/3}$, which is true for $E \rightarrow \infty$, matches onto the formula given above. This is completely expected, since in the case of $d=2$ the bound $\rho(\Delta) \lesssim e^{2\pi \Delta}$ implied by the phase transition assumed here can be used to prove the validity of the Cardy formula down to $\Delta \sim c/6$  \cite{Hartman:2014oaa}. So this result is weaker, but the different route taken is illuminating and can potentially be applied in other cases where arguments like that of \cite{Hartman:2014oaa} are absent. 

\section{AdS$_5$, AdS$_6$, and AdS$_7$ details}\label{higherdetails}
In section \ref{genericsec}, we exhibited results for bounds on the density of states with $d=4,5,$ and $6$ dimensional boundaries. The metrics and derivation of thermodynamic quantities, including the Euclidean actions are straightforward and follow the same steps as in $d=2,3$ but the expressions are longer and not directly illuminating. Below, we expound on the steps that lead to the bounds above. In particular, we collect results for AdS$_6$ and AdS$_7$ whose derivation is distributed over multiple papers in the literature. First, we discuss the derivation for $d=4$. We set $G=1$ everywhere below.
\subsection{AdS$_5$}\label{AdS5}
Here we follow \cite{Chong:2005hr}. The relevant thermodynamic quantities are
\begin{align}
\Omega_a &= \frac{a(r_+^2+b^2)(1+r_+^2)+bq}{(r_+^2+a^2)(r_+^2+b^2)+abq}, \qquad \Omega_b = \Omega_a(a\leftrightarrow b), \qquad \Phi =\frac{\sqrt{3}qr_+^2}{(r_+^2+a^2)(r_+^2+b^2)+abq}, \nonumber\\ 
J_a &= \frac{\pi[2am+qb(1+a^2)]}{4\Xi_a^2\Xi_b}, \qquad J_b = J_a(a\leftrightarrow b),\qquad Q=\frac{\sqrt{3}\pi q}{4\Xi_a\Xi_b}\nonumber\\ 
M &= \pi\frac{m[2(\Xi_a+\Xi_b)-\Xi_a\Xi_b]+2qab(\Xi_a+\Xi_b)]}{4\Xi_a^2\Xi_b^2},\qquad \beta = \frac{2\pi r_+[(r_+^2+a^2)(r_+^2+b^2)+abq]}{r_+^4[1+2r_+^2+a^2+b^2]-(ab+q)^2}\nonumber
\end{align}
where $\Xi_a = 1-a^2, \Xi_b = 1-b^2$ and $m = [(r^2+a^2)(r^2+b^2)(1+r^2)+q^2+2abq]/2r^2$. The vacuum subtracted Euclidean action is
\begin{align}
\Delta I_E = \frac{\pi\beta}{8\Xi_a\Xi_b r_+^2}\left[(r_+^2+a^2)(r_+^2+b^2)(1-r_+^2)+2abq+q^2\left(1-\frac{2r_+^4}{(r_+^2+a^2)(r_+^2+b^2)+abq}\right)\right].
\end{align}
Analytic results are possible for $q=0$ or $b=0$; however, numerics are necessary in the generic case.  The BPS limit of these black holes is $E=|J_1|+|J_2|+\sqrt{3}|Q|$ beyond which there are no states. Note, this bound differs slightly from the Chamblin et al. case (see \ref{chargebound}), because of a factor of 2 in the definition of the Maxwell field.

\subsection{AdS$_6$}\label{d5section}
\label{AdS6}

The bounds on the density of states are meant to serve all holographic CFTs in their respective dimensions. However, there are no bottom-up solutions for Einstein-Maxwell gravity in $d=5$ and $d=6$. This may not be surprising as higher form fields and Chern-Simons terms seem natural in higher dimensions, especially in consistent supergravity trunctations. Instead, one must search for the most generic supergravity solution with AdS asymptotics and fewest bulk fields. The most generic choice we could find in the literature is the black hole in \cite{Chow:2008ip}. This comes from a dimensional reduction of massive type IIA supergravity on a hemisphere of $S^4$ \cite{Cvetic:1999un}. This supergravity theory should arise as the near horizon limit of a D4-D8 brane configuration and is dual to a $d=5, \;\mathcal{N}=2$ superconformal field theory. The bosonic field content of six dimensional $\mathcal{N}=4, SU(2)$ gauged supergravity is a graviton, a two-form potential, a one-form potential, the gauge potentials of $SU(2)$ Yang-Mills and a scalar. We can truncate to the sector where only one $U(1)$ of the SU(2) is excited. Then, the bosonic Lagrangian is
\begin{align}
\mathcal{L} = R\star 1-\frac{1}{2}\star d\phi\wedge d\phi - X^{-2}\left(\star F_{(2)}\wedge F_{(2)}+g^2\star A_{(2)}\wedge A_{(2)}\right) - \frac{1}{2}X^4\star F_{(3)}\wedge F_{(3)}\nonumber\\
+g^2\left(9X^2+12X^{-2}-X^{-6}\right)\star 1 - F_{(2)}\wedge F_{(2)}\wedge A_{(2)} - \frac{g^2}{3}A_{(2)}\wedge A_{(2)}\wedge A_{(2)},
\end{align}
with $F_{(2)}=dA_{(1)}$. We now set $g=1$.

The relevant thermodynamic quantities are
\begin{align}
\Omega_a&=\frac{a[(1+r_+^2)(r_+^2+b^2)+qr_+]}{(r_+^2+a^2)(r_+^2+b^2)+qr_+}, \qquad \Omega_b=\Omega_a(a\leftrightarrow b), \qquad \Phi = \frac{\sqrt{q(2m+q)}r_+}{(r_+^2+a^2)(r_+^2+b^2)+qr_+}\\ 
J_a &=\frac{\pi a(2m+\Xi_bq)}{3\Xi_a^2\Xi_b}, \qquad J_b = J_a(a\leftrightarrow b), \qquad Q=\frac{\sqrt{q(2m+q)}}{\Xi_a\Xi_b}\nonumber\\
M &=\frac{\pi}{3\Xi_a\Xi_b}\left[2m\left(\frac{1}{\Xi_a}+\frac{1}{\Xi_b}\right)+q\left(1+\frac{\Xi_a}{\Xi_b}+\frac{\Xi_b}{\Xi_a}\right)\right],\quad S=\frac{2\pi^2\left[(r_+^2+a^2)(r_+^2+b^2)+qr_+\right]}{3\Xi_a\Xi_b}\nonumber\\
\beta&=\frac{4\pi r_+[(r_+^2+a^2)(r_+^2+b^2)+qr_+]}{2(1+r_+^2)r_+^2(2r_+^2+a^2+b^2)-(1-r_+^2)(r_+^2+a^2)(r_+^2+b^2)+4qr_+^3-q^2}.
\end{align}
where $\Xi_a = 1-a^2, \Xi_b = 1-b^2$ and
\begin{align}
m=\frac{(r^2+a^2)(r^2+b^2)+[r(r^2+a^2)+q][r(r^2+b^2)+q]}{2r}.
\end{align}
The Gibbs free energy, defined by 
\begin{align}
G= E-TS-\Phi Q-J_a\Omega_a-J_b\Omega_b
\end{align}
is equivalent to the background subtracted on-shell Euclidean action divided by $-\beta$, $\Delta I_E=-\beta G$. Plugging everything in, we get
\begin{align}
\Delta I_E &= \frac{\pi \beta}{6 r_+ \Xi _a \Xi _b \left(r_+^2 \left(a^2+b^2\right)+a^2 b^2+q r_++r_+^4\right)}\biggl[q^2 \left(-r_+^2 \left(a^2+b^2\right)+a^2 b^2-3 r_+^4\right)\nonumber\\&-q r_+ \left(3 r_+^2-1\right) \left(a^2+r_+^2\right) \left(b^2+r_+^2\right)-\left(r_+^2-1\right) \left(a^2+r_+^2\right){}^2 \left(b^2+r_+^2\right){}^2-q^3 r_+\biggr].
\end{align}
In the limit of zero charge, this agrees with the generically spinning black holes in six dimensions with no charge \cite{Gibbons:2004uw}. However, it turns out the $a=b=0$ solution is not the AdS-Reissner-Nordstr\"om black hole, but rather the black hole in \cite{Cvetic:1999un}. In table \ref{analyticdensities}, for the charged static case, we instead presented the result from \cite{Chamblin:1999tk}, where the action is the one calculated in section \ref{chargedanalytic}.


\subsection{AdS$_7$}
\label{AdS7}

The $d=6$ case follows \cite{Chow:2007ts}. These solutions come from reducing eleven-dimensional supergravity on $S^4$ leading to seven dimensional $\mathcal{N}=4, SO(5)$ gauged supergravity. Note that this can be thought of as coming from the near horizon limit of a stack of $M5$ branes and is dual to the six-dimensional, $\mathcal{N}=(2,0)$ SCFT. For singly charged black holes, we choose to truncate to the $U(1)^3$ Cartan subgroup. The bosonic fields are a graviton, a self dual 3-form potential, two $U(1)$ gauge fields and two scalars. Turning off one of the scalars in the gauged theory sets the two $U(1)$ fields equal and the Lagrangian is
\begin{align}
\mathcal{L} = &R\star 1 - \frac{1}{2}\star d\phi_1\wedge d\phi_1 -X^{-2}\star F_{(2)}\wedge F_{(2)}-\frac{1}{2}X^4 \star F_{(4)}\wedge F_{(4)}\nonumber\\ &+2g^2(8X^2+8X^{-3}-X^{-8})\star 1 
+ F_{(2)}\wedge F_{(2)}\wedge A_{(3)} -g  F_{(4)}\wedge A_{(3)},
\end{align}
where $X=e^{-\phi_1/\sqrt{10}}$. The self-duality condition reads
\begin{align}
X^4\star F_{(4)} = 2gA_{(3)}-dA_{(2)}+F_{(2)}\wedge A_{(1)}.
\end{align}
For this work, we set $g=-1$ (this must be negative for BPS states). As in six dimensions, it is more straightforward to calculate the Gibbs free energy. The relevant thermodynamic quantities are
\begin{align}
\Omega_i &= \frac{a_i[(1+r_+^2)\prod_{j\neq i}(r_+^2+a_j^2)+qr_+^2]+q\prod_{j\neq i}a_j}{\prod_i(r_+^2+a_i^2)+q(r_+^2+abc)},\qquad \Phi = \frac{\sqrt{q(2m+q)} r_+^2}{\prod_i(r_+^2+a_i^2)+q(r_+^2+abc)},\nonumber\\
J_i &= \frac{\pi^2[a_i(2m+q)+q(\Pi_{j\neq i}a_j-a_i\sum_{j\neq i}a_j^2 +abca_i)]}{8\Xi_a\Xi_b\Xi_c\Xi_i},\qquad Q = \frac{\pi^2m\sqrt{q(2m+q)}}{2\Xi_a\Xi_b\Xi_c}, \nonumber \\
E &=\frac{\pi^2}{8\Xi_a\Xi_b\Xi_c}\left[\sum_i\frac{2m}{\Xi_i}-m+\frac{5q}{2}+\frac{q}{2}\sum_i\left(\sum_{j\neq i}\frac{2\Xi_j}{\Xi_i}-\Xi_i-\frac{2(1-2abc)}{\Xi_i}\right)\right]\nonumber \\
\beta&=\frac{2\pi r_+[\prod_i(r_+^2+a_i^2)+q(r_+^2+abc)]}{(1+r_+^2)r_+^2\sum_i\prod_{j\neq i}(r_+^2+a_i^2)-\prod_i(r_+^2+a_i^2)+2q(r_+^4-abc)-q^2} \nonumber.
\end{align}
For brevity, we used $a_i\in \{a,b,c\}$. The parameter $m$ is
\begin{align}
2m&=\frac{1+r^2}{r^2}(r^2+a^2)(r^2+b^2)(r^2+c^2)+q(2r^2+a^2+b^2+c^2)+\frac{2qabc}{r^2}+\frac{q^2}{r^2}.
\end{align}
Now the regularized Euclidean action is
\begin{align}
\Delta I_E&=\frac{\beta\pi^2}{16\Xi_a\Xi_b\Xi_c}\Biggl[(1-r_+^2)\prod_i(r_+^2+a_i^2)-2qr_+^4+2qabc\nonumber\\
&-q^2\left(\sum_ia_i^2r_+^4-\sum_{i<j}a_i^2a_j^2r_+^2-\prod_ia_i^2-abc(2r_+^4-2r_+^2+q)+r_+^2(r_+^4+q)\right)(\prod_i(r_+^2+a_i^2)+q(r_+^2+abc))^{-1}\Biggr].
\end{align}
The limit $q\to0$ agrees with the Myers-Perry-AdS$_7$ solutions, but like $d=5$, the non-spinning limit does not match the Reissner-Nordstr\"om result of Chamblin et al.

\footnotesize
\bibliographystyle{JHEP}
\bibliography{local}

\end{document}